%% file: habil.tex
\documentclass[twoside]{report}
\usepackage{epsf,amssymb,amsmath}

\renewcommand{\ln}{ {\rm ln} }

\newcommand{\urbsix}{\cite{urb96}}
\newcommand{\urbseven} {\cite{urb97}}
\newcommand{\urbsevenb}{\cite{urb97b}}
\newcommand{\schrseven}{\cite{schr97}}

\newcommand{\R}{{\mathbb R}}

\newcommand{\G}{ {\cal N} }
\newcommand{\D}{{\mathbb D}}
\newcommand{\sign}{ \mbox{\rm sgn} }
\newcommand{\ext}{ \mbox{\rm extr} }
\newcommand{\exta}[1]
     {{\renewcommand{\arraystretch}{0.75} \begin{array}[t]{c} 
       \ext \\ {\scriptstyle #1}
     \end{array}}}

\newcommand{\half}{{\frac{1}{2}}}
\renewcommand{\#}{\displaystyle}

\newcommand{\La}{\left\langle}
\newcommand{\Ra}{\right\rangle}
\newcommand{\cut}[1]{}

\begin{document}
\baselineskip 1.1\baselineskip

\ \\[2cm]
\begin{center} {\Huge\bf
Statistical Physics of \\[0.3cm]
Feedforward Neural Networks}\\[5cm]

{\LARGE Robert Urbanczik}\\[1cm]

{\large 
 Institut f\"ur theoretische Physik \\
 Universit\"at W\"urzburg \\
 Am Hubland \\
 D-97074 W\"urzburg \\
 Germany }

\vspace*{4cm}
\Large W\"urzburg, January 2002
\end{center} 
\vfill
\pagestyle{empty}
\pagebreak\ \pagebreak

\noindent{\Huge\bf Abstract}\\[1cm]

{\large \noindent
The article is a lightly edited version of my habilitation thesis
at the University W\"urzburg. My aim is to give a self contained,
if concise, introduction to the formal methods used when off-line
learning in feedforward networks is analyzed by statistical physics.
However, due to its origin, the article is not a comprehensive
review of the field but is highly skewed towards reporting my own 
research.}

\vfill
\pagestyle{empty}
\pagebreak\ \pagebreak
 
\noindent{\Huge\bf Preface}\\[1cm]

\noindent
This thesis summarizes my postdoctoral research in sofar as it dealt with
supervised learning in feedforward neural networks. This research was carried
out at the University of W\"urzburg and at Aston University, Birmingham.

In  W\"urzburg, I wish to thank Wolfgang Kinzel for giving me the opportunity
to work in the stimulating atmosphere of his group. I also acknowledge
many interesting discussions with Georg Reents and Michael Biehl. Michael
deserves special thanks for taking trouble to comment on the introductory
chapters of this work. In Birmingham my thanks go to David Saad and 
Manfred Opper, both for fruitful discussions and for doing their best to make
my stay in England more enjoyable.

Obviously, there are many more people I need to thank - I have not even 
mentioned all of my coauthors. As a catch all, let me extend my warm thanks to 
everyone I am omitting to thank. I do, however, wish to explicitly 
acknowledge the support of the {\em Deutsche Forschungsgemeinschaft} which 
funded a substantial part of the research reported below.

\vfill
\pagestyle{empty}
\pagebreak\ \pagebreak

\renewcommand{\thepage}{}
\tableofcontents

\vfill\pagebreak\ \pagebreak

\setcounter{page}{1}
\renewcommand{\thepage}{\arabic{page}} 
\pagestyle{headings}

\include{intro}

\include{gardner}

\include{extram}

\include{mlp}

\include{mlp2}

\pagebreak
\begin{appendix}

\newcommand{\Underline}[1]{ {#1} }

\chapter{The entropy term}

\noindent
We want to calculate a volume of the form 
\begin{equation}
D_n({\mathbf{\mathsf{Q}}}) = \int d{\mathbf{\mathsf{J}}} \,\, \mbox{\boldmath$\delta$}
({\mathbf{\mathsf{Q}}} -  
{\mathbf{\mathsf{J}}}^\top{\mathbf{\mathsf{J}}}) = 
\int d{\mathbf{\mathsf{J}}} \prod_{a,b=1 (a\leq b)}^n \delta\left(\, Q^{ab}- 
{\Underline{J}^a}^T  \Underline{J}^b \,\right)
\end{equation}
 where $\mathbf{\mathsf{Q}}$ is
a symmetric, positive definite $(n,n)$-matrix of overlaps and 
$\mathbf{\mathsf{J}}$ is the $(N,n)$-matrix which
is composed of the $n$ vectors $\Underline{J}^a \in \mbox{I}\!\mbox{R}^N$.

For a suitable orthogonal $(n,n)$-matrix $\mathbf{\mathsf{o}}$ and a diagonal $(n,n)$-matrix
$\mathbf{\mathsf{D}}$ one can write $\mathbf{\mathsf{Q}}$ as 
$\mathbf{\mathsf{Q}} = \mathbf{\mathsf{o}}^\top\mathbf{\mathsf{DDo}}$. We now apply the linear
transformation $\mathbf{\mathsf{J}} \rightarrow \mathbf{\mathsf{JDo}}$ 
to the above integral.
Its determinant is $\det \mathbf{\mathsf{D}}^N$ and we obtain 
\begin{eqnarray}
D_n(\mathbf{\mathsf{Q}})
     &=& \int d{\mathbf{\mathsf{J}}} \,\,\, {\mbox{\boldmath$\delta$}} 
     ({\mathbf{\mathsf{o}}}^\top{\mathbf{\mathsf{D}}}({\mathbf{\mathsf{1}}}-
      {\mathbf{\mathsf{J}}}^\top{\mathbf{\mathsf{J}}}){\mathbf{\mathsf{Do}}}) 
     \,\,\, \det {\mathbf{\mathsf{D}}}^N\;.  \label{transJ}
\end{eqnarray}
The Fourier representation of the $\delta$-function yields
\begin{equation}
{\mbox{\boldmath$\delta$}}({\mathbf{\mathsf{o}}}^\top {\mathbf{\mathsf{D}}}
({\mathbf{\mathsf{1}}}-{\mathbf{\mathsf{J}}}^\top{\mathbf{\mathsf{J}}})
 {\mathbf{\mathsf{Do}}}) = C_n \int d\hat{{\mathbf{\mathsf{Q}}}} 
 \,\, \exp\left(\, i \, \mbox{\rm Tr} \left[ \hat{{\mathbf{\mathsf{Q}}}}
{\mathbf{\mathsf{o}}}^\top{\mathbf{\mathsf{D}}}({\mathbf{\mathsf{1}}}-
{\mathbf{\mathsf{J}}}^\top{\mathbf{\mathsf{J}}}){\mathbf{\mathsf{Do}}}\right] \right)
\,.
\end{equation}
The integration runs over symmetric $(n,n)$-matrices and 
$C_n = (2\pi)^{-n(n+1)/2}2^{n(n-1)/2}$, where the second factor arises
from the fact that the off-diagonal elements are counted twice in the
trace.
Using  
$$ \mbox{Tr}\left[\,\hat{{\mathbf{\mathsf{Q}}}}{\mathbf{\mathsf{o}}}^\top
{\mathbf{\mathsf{D}}}({\mathbf{\mathsf{1}}}-{\mathbf{\mathsf{J}}}^\top
{\mathbf{\mathsf{J}}}){\mathbf{\mathsf{Do}}}\, \right] \,\,  = \,\,  \mbox{Tr}\left[\,
{\mathbf{\mathsf{Do}}}\hat{{\mathbf{\mathsf{Q}}}}{\mathbf{\mathsf{o}}}^\top
{\mathbf{\mathsf{D}}}({\mathbf{\mathsf{1}}}-
{\mathbf{\mathsf{J}}}^\top {\mathbf{\mathsf{J}}}) \, \right]$$ 
and 
transforming 
$\hat{{\mathbf{\mathsf{Q}}}}$ via 
$\hat{{\mathbf{\mathsf{Q}}}} \to  {\mathbf{\mathsf{o}}}^\top
{\mathbf{\mathsf{D}}}^{-1} \hat{{\mathbf{\mathsf{Q}}}}
{\mathbf{\mathsf{D}}}^{-1} {\mathbf{\mathsf{o}}} $ yields

\begin{eqnarray} {\mbox{\boldmath$\delta$}}
( {\mathbf{\mathsf{o}}}^\top{\mathbf{\mathsf{D}}} ({\mathbf{\mathsf{1}}}-
{\mathbf{\mathsf{J}}}^\top{\mathbf{\mathsf{J}}}
{\mathbf{\mathsf{Do}}}) & =  & 
C_n \det {\mathbf{\mathsf{D}}}^{-n-1} \int \, d\hat{{\mathbf{\mathsf{Q}}}} 
\,\, \exp\left(\,i\, \mbox{Tr} \left[
\hat{{\mathbf{\mathsf{Q}}}}(
{\mathbf{\mathsf{1}}}-
{\mathbf{\mathsf{J}}}^\top{\mathbf{\mathsf{J}}})
\right])\right) \nonumber \\
                  & = & C_n \det {\mathbf{\mathsf{D}}}^{-n-1} 
{\mbox{\boldmath$\delta$}}
({\mathbf{\mathsf{1}}}-{\mathbf{\mathsf{J}}}^\top{\mathbf{\mathsf{J}}})
\end{eqnarray}
and thus $D_n({\mathbf{\mathsf{Q}}}) = \det {\mathbf{\mathsf{D}}}^{N-n-1} 
D_n({\mathbf{\mathsf{1}}})$.
Of course 
$\det {\mathbf{\mathsf{D}}}^2 = \det {\mathbf{\mathsf{Q}}}$, so  finally
\begin{equation}
D_n(\mathbf{\mathsf{Q}}) = 
D_n({\mathbf{\mathsf 1}})(\det \mathbf{\mathsf{Q}})^{(N-n-1)/2}
\end{equation}
where $D_n({\mathbf{\mathsf{1}}})$ is just a normalization
constant.

The case where one considers an additional $(N,m)$-Matrix ${\mathbf{\mathsf{B}}}$
of $m$ teacher vectors
and wants to evaluate   $\int \, d{\mathbf{\mathsf{J}}}\,\,
{\mbox{\boldmath$\delta$}} ( 
{\mathbf{\mathsf{Q}}} - 
{\mathbf{\mathsf{J}}}^\top
{\mathbf{\mathsf{J}}}) 
\, {\mbox{\boldmath$\delta$}} ( 
{\mathbf{\mathsf{R}}}-{\mathbf{\mathsf{J}}}^\top 
{\mathbf{\mathsf{B}}}) 
\, $ reduces to the above consideration by noting that the integral will 
not depend on the choice of 
${\mathbf{\mathsf{B}}}$, as long as the matrix of teacher overlaps
${\mathbf{\mathsf{T}}}= {\mathbf{\mathsf{B}}}^\top{\mathbf{\mathsf{B}}} $
is held fixed. Thus, one may in addition integrate over 
all ${\mathbf{\mathsf{B}}}$ which have correlation matrix ${\mathbf{\mathsf{T}}}$.
\end{appendix}

\addcontentsline{toc}{chapter}{Bibliography}
\bibliographystyle{plain}
\bibliography{../tex/neural} 
\end{document}

%% file: gardner.tex
\chapter{Capacity of the perceptron}

Choosing a weight vector $J \in \R^N$ defines a dichotomy of the $P$
inputs $\xi^\mu \in \R^N$ by classifying an input as  $\sign(J^T\xi^\mu)$.
We now calculate the number $C(P,N)$ of dichotomies which can be 
obtained in this manner by an inductive argument due to  Schl\"afli.%
\footnote{To simplify the argument, we only count the dichotomies which
can be obtained with a $J$ satisfying $J^T\xi^\mu \neq 0$ for all $\mu$.}
Let $\xi$ be an additional input  and assume that all points are in general
position. Let $D$ be the number of dichotomies on $\xi^1,\ldots,\xi^P$
which can be represented by a weight vector $J$ satisfying $J^T\xi=0$.
For any such dichotomy we obtain two dichotomies differing only on $\xi$ by
replacing $J$ with $J'=J\pm\epsilon\xi$ and choosing $\epsilon$ sufficiently
small. Hence $C(P+1,N) = C(P,N) + D.$ Further $J^T\xi=0$, the constraint 
defining $D$, means that $J$ is confined to an $N-1$ dimensional subspace,
so $D = C(P,N-1)$. Finally, the recursion
\begin{equation*}
C(P+1,N) = C(P,N) + C(P,N-1)
\end{equation*} 
with boundary conditions $C(1,N) = C(P,1) = 2$ has the solution
\begin{equation*}
C(P,N) = 2 \sum_{i=0}^{N-1} \binom{p-1}{i}\,.
\end{equation*}

In particular the perceptron can implement all possible dichotomies,
$C(P,N) = 2^P$, only if $P\leq N$. But considering large $N$ and scaling
$P$ as $P = \alpha N$ one finds
\begin{equation}
\lim_{N\rightarrow\infty} 
\frac{C(\alpha N,N)}{2^{\alpha N}} = 
  \left\{ \begin{array}{cl} 1 &\mbox{if } \alpha < 2 \\
                            0 &\mbox{if } \alpha > 2\,.
          \end{array}
  \right. 
\label{Schlaef}
\end{equation}

So in the limit of large $N$ almost all possible dichotomies can be 
implemented as long as $P/N < 2$.

The above result was rederived by Elisabeth Gardner
using the very different approach of Statistical Physics.
While Gardner's calculation is more involved, it can
be adapted to many related  scenarios and in particular is a starting point 
for analyzing learning in multilayer networks.

For a given dichotomy, let $\tau^\mu \in \{-1,1\}$ be the labels of the
inputs $\xi^\mu$ and we shall call the input/output pairs 
$\D = \{ (\xi^\mu,\tau^\mu) \}_{\mu=1}^P$ the training set.
A perceptron with weight vector $J$ implements
the dichotomy if $\tau^\mu J^T \xi^\mu > 0$ for all $\mu$ and for convenience 
we may assume that the Euclidian norm of $J$ equals $1$. In terms of the
Heaviside step function 
\begin{equation*}
\Theta(x) = \left\{ \begin{array}{cl} 0 &{\rm if}\ x \leq 0 \\
                                      1 &{\rm if}\ x   > 0 
                    \end{array} \right .
\end{equation*}
the volume $V(\D)$  of the
weight vectors implementing the dichotomy can then be written as
\begin{equation}
V(\D) = \int \!{\rm d}J \prod_{\mu=1}^P \Theta(\tau^\mu J^T\xi^\mu)\,.
\label{VD}
\end{equation}
The integration is over the unit sphere in $\R^N$ and we normalize the
measure such that $V(\emptyset)=1$.

In statistical mechanics one is interested in the properties of $V(\D)$
given a distribution of training sets $\D$. We shall always assume
that the patterns $(\xi^\mu,\tau^\mu)$ in a training set are obtained
by {\em independently} sampling a random variable $(\xi,\tau)$ with
values in $\R^N\times \{-1,1\}$. For the storage capacity problem considered
by Gardner $\tau$ is further assumed independent of $\xi$, and
$\tau^\mu = \pm 1$ with equal probability. When averaging over all
training sets $\D$ of size $P$ one then obtains 
$\La V(\D) \Ra_{\D} =\La V(\{(\xi,\tau)\}) \Ra_{(\xi,\tau)}^P = 2^{-P}$.

Despite its simplicity this result is remarkable when compared to
Eq. (\ref{Schlaef}). For large $N$ and $P > 2 N$ Eq. (\ref{Schlaef}) means
that for almost all training sets $V(\D) = 0$. This is not at all reflected
in the behavior of  $\La V(\D) \Ra_{\D}$; so there must be a few
training sets for which $V(\D)$ is very large compared to  $2^{-P}$.
Instead of averaging $V(\D)$, one thus has to consider quantities
such as $\La \Theta(V(\D)) \Ra_{\D}$, the probability that a dichotomy
can be implemented, or $\La \ln V(\D) \Ra_{\D}$, which will diverge
if the probability that $V(\D) = 0$ is finite. Calculating these averages
analytically is, however, quite difficult, but they could easily be obtained
if one knew $\La V^n(\D) \Ra_{\D}$ for all real $n$.

The basic idea of the replica method is to calculate the moments of $V(\D)$,
that is to consider $\La V^n(\D) \Ra_{\D}$ for 
the special case that $n$ is a natural number. In contrast
to general $n\in \R$, this case is tractable and it turns out that
the expression $g(n)$ for the $n$-th moment thus found 
can be evaluated for real $n$ and is even an analytical function of $n$.
So assuming this analytical continuation to be correct, i.e.
$\La V^n(\D) \Ra_{\D} = g(n)$ for all positive $n$,
one then for instance obtains the probability that a dichotomy can be 
implemented as $\La \Theta(V(\D)) \Ra_{\D} = \lim_{n\rightarrow+0} g(n)$.

\newcommand{\dd}[1]{ {\rm d} #1 }
\newcommand{\J}{ {\mathsf J} } 

The replica method is not just applicable when the random variable in
question is a volume and we shall straight away consider the more general 
form
\begin{equation}
Z(\D)  = \int \!{\rm d}J \prod_{\mu=1}^P F(\tau^\mu J^T\xi^\mu)\,;
\label{part}
\end{equation}
so $Z(\D) = V(\D)$ for the special case that $F$ is the $\Theta$-function.
We assume that $F$ is nonnegative and that the RHS of Eq. (\ref{part}) as well
as some related integrals are well defined. 
The name replica method is motivated by the fact that $Z^n(\D)$ is an $n$-fold
integral for integer $n$:
\begin{equation*}
Z^n(\D) = \int \!\dd{\J} \prod_{\mu=1}^P \prod_{a=1}^n
           F(\tau^\mu {J^a}^T\xi^\mu)\,,
\end{equation*}
where $\dd{\J} \equiv \prod_a {\rm d}J^a$. For the moments of 
$Z(\D)$ one then has
\begin{equation}
\La Z^n(\D) \Ra_\D  = \int \!\dd{\J} \La  \prod_{a=1}^n  
                       F(\tau {J^a}^T\xi) \Ra_{(\xi,\tau)}^P \,,
\label{Grep}
\end{equation}
since the examples are independent.
To evaluate the average one has to make some assumptions about the distribution
of the inputs and it is simplest to assume that the components
$\xi_i$ of $\xi$ are i.i.d. $\G(0,1)$, that is independent and Gaussian with
zero mean and unit variance. Then the distribution of
the inner products ${J^a}^T\xi$ in Eq. (\ref{Grep}) is Gaussian as well,
with zero
mean and covariances $\La {J^a}^T\xi {J^b}^T\xi \Ra_{\xi} = {J^a}^T J^b$.
Consequently if  $X(Q)$ is an $n$-dimensional Gaussian of zero mean and
with a covariance matrix $Q$ satisfying $Q_{ab} = {J^a}^T J^b$ one has
\begin{equation}
\La  \prod_{a=1}^n F(\tau {J^a}^T\xi) \Ra_{(\xi,\tau)} =
\La  \prod_{a=1}^n F(\tau X_a(Q) ) \Ra_{X(Q),\tau}     =
\La  \prod_{a=1}^n F( X_a(Q) ) \Ra_{X(Q)},
\label{Ggauss}
\end{equation}
where the last equality holds because $X(Q)$ and $-X(Q)$ have the same 
distribution. Since the integrand in Eq. (\ref{Grep}) depends on the weight 
vectors $J^a$ only via their overlaps ${J^a}^T J^b$, it is convenient to 
transform the integration variables. This is best done by multiplying
Eq. (\ref{Grep}) with 
\begin{equation}
\int \dd{Q}\, \delta( Q - \J^T \J) :=  
    \int \dd{Q} \prod_{a < b \leq n} \delta( Q_{ab} - {J^a}^TJ^b ) = 1
\label{Gone}
\end{equation}
and changing the order of integration. The integral over $Q$ runs
over the symmetric and positive definite matrices with $Q_{aa}=1$. 
Combining Eqs. 
(\ref{Grep},\ref{Ggauss},\ref{Gone}) then yields
\begin{equation*}
\La Z^n(\D) \Ra_\D  = \int \!\dd{Q}  D_n(Q)
\La  \prod_{a=1}^n F( X_a(Q) ) \Ra_{X(Q)}^P \,,
\end{equation*}
where $D_n(Q) = \int \!\dd{\J}\, \delta( Q - \J^T \J)$. In the appendix
a simple derivation is given
that $D_n(Q) = D_n({\mathsf 1})(\det Q)^{(N-n-1)/2}$ where ${\mathsf 1}$ is 
the $n$ by $n$ identity matrix. Thus, setting $P=\alpha N$,
\begin{equation*}
\La Z^n(\D) \Ra_\D  = D_n({\mathsf 1}) \int \!\dd{Q} 
\left(  (\det Q)^\frac{1-(n+1)/N}{2}
\La  \prod_{a=1}^n F( X_a(Q) ) \Ra_{\!\!X(Q)}^\alpha 
\right )^N . 
\end{equation*}
Now the integration is over $n(n-1)/2$ dimensions and this number does
not increase
with $N$. We may thus use that the $L_N$-norm converges to the maximum
norm with increasing $N$ to find
\begin{equation}
\lim_{N\rightarrow\infty}N^{-1} \ln \La Z^n(\D) \Ra_\D  = 
\max_Q\, \half\ln\det Q +
\alpha \ln \La  \prod_{a=1}^n F( X_a(Q) ) \Ra_{X(Q)}. 
\label{Glab}
\end{equation}

Solving this extremal problem for general $Q$ is quite difficult and one
thus restricts the search to a small subspace  of all possible
matrices $Q$. If one assumes that the extremal problem has a unique solution
$Q^*$, all off diagonal elements of $Q^*$ must have the same value since
the set of solutions is invariant under permutations of the replica indices.
This is known as the replica symmetric assumption.

One is thus lead to consider $n$ by $n$ matrices $M_n(u,v)$ with diagonal
elements equal to $u$ and off diagonal elements equal to $v$. A simple 
calculation shows that $(1,1,\ldots,1)^T$ is an eigenvector of $M_n(u,v)$
with eigenvalue $u+(n-1)v$ and that the matrix further has
$n-1$ linearly independent eigenvectors of the form 
$(1,0,\ldots,0,-1, 0,\ldots,0)^T$ with eigenvalue $u-v$. Thus
$\det M_n(u,v) = (u+(n-1)v)(u-v)^{n-1}$. Assuming replica symmetry,
$Q^* = M_n(1,q)$, and we can simplify the first term in Eq. (\ref{Glab}).

Further if the $n+1$
random variables $z_0,\ldots,z_n$ are i.i.d. $\G(0,1)$, the covariance
matrix of the linear combinations
$X_a = \sqrt{u^2-v^2} z_a + v z_0$, $a=1,\ldots,n$,
is just $M_n(u^2,v^2)$. 
For $Q^* = M(1,q)$ this  
observation enables us to factorize the average in Eq. (\ref{Glab}) as
\begin{eqnarray*}
 \La  \prod_{a=1}^n F( X_a(Q^*) ) \Ra_{X(Q^*)} &=&
 \La  \prod_{a=1}^n F(\sqrt{1-q} z_a + \sqrt q z_0  ) \Ra_{z_0,\ldots,z_n} 
  \\ &=&
 \La \La F(\sqrt{1-q} z_1 + \sqrt q z_0  )\Ra_{z_1}^n \Ra_{z_0}.
\end{eqnarray*}
So in replica symmetry we obtain the important intermediate
result
\begin{equation}
\lim_{N\rightarrow\infty}N^{-1} \ln \La Z^n(\D) \Ra_\D  = 
\max_q\, f(n,q)\,,
\end{equation}
where 
\begin{equation}
f(n,q) = \half\ln(1+(n-1)q) +\frac{n-1}{2}\ln(1-q) + 
\alpha\ln \La  \La F(\sqrt{1-q} z_1 + \sqrt q z_0  )\Ra_{z_1}^n   \Ra_{z_0}. 
\label{Gfunc}
\end{equation}
Now $f(n,q)$ is well defined for any nonnegative value of $n$ and
not just for integer $n$. We can thus consider the analytical continuation
to
values of $n$ close to zero. Here, one has to be rather careful since
$f(n,q)$ does not depend on $q$ when $n=1$, and in particular
$f(1,q) = -\alpha\ln \La F(z_0)\Ra_{z_0}$.
Expanding
$f(n,q)$ around $n=1$ thus yields
\begin{equation*}
f(n,q) \approx -\alpha\ln\La F(z_0)\Ra_{z_0}  + (n-1)f_1(1,q)\,
\end{equation*}
where $f_1$ is the partial derivative of $f$ w.r.t to the first argument.
So if $q^*$ maximizes $f_1(1,q)$ and if $n$ is greater but close to
$1$,
$f(n,q^*)$ will be a good approximation to the maximum of $f(n,q)$. 
But by the
same argument $f(n,q^*)$ will be a good approximation to the {\em minimum}
of $f(n,q)$ when $n$ is close to but smaller than $1$. When looking
for a function $q(n)$ such that $f(n,q(n))$ is analytical, one will at 
least
want $q(n)$ to be continuous at $n=1$. Hence  $f(n,q)$ must be {\em minimized} 
when $n<1$. So for small positive values of $n$ we obtain
\begin{equation}
\lim_{N\rightarrow\infty}N^{-1} \ln \La Z^n(\D) \Ra_\D  = 
{\cal O}(n^2) + n \min_q\, f_1(0,q)\,,
\label{Gmin}
\end{equation} 
and from Eq. (\ref{Gfunc})
\begin{equation*}
 f_1(0,q) = \half \left( \frac{q}{1-q} + \ln(1-q) \right) +
     \alpha \La \ln\La F(\sqrt{1-q} z_1 + \sqrt q z_0  )\Ra_{z_1} \Ra_{z_0}\,.
\end{equation*}
The first term in the above sum is often called the {\em entropy term} since
it is determined by the constraints on the weight vectors of the perceptron. 
In the present case, continuous and normalized weight vectors. The second term,
which depends on the choice of $F$, is called the {\em energy term}.

We now specialize to Gardner's case where $Z(\D)$ is the volume
$V(\D)$ and $F(x) = \Theta(x)$. Introducing the function 
$H(x) = \La \Theta(z_1-x) \Ra_{z_1}$ which is closely related to
the error function, the expression for $f_1$ simplifies to
\begin{equation}
 f_1(0,q) = \half \left( \frac{q}{1-q} + \ln(1-q) \right) +
            \alpha \La \ln H(-z_0 \sqrt q/\sqrt{1-q}\,)  \Ra_{z_0}\,.
\label{Glog}
\end{equation}

Since the probability that a dichotomy can be implemented by the perceptron
is
$\La \Theta(V(\D)) \Ra_{\D} = 
  \lim_{n\rightarrow+0} \ln \La V^n(\D) \Ra_\D$,
by Eq. (\ref{Gmin})
this probability approaches  $1$ in the limit of large $N$ if 
$\min_q\, f_1(0,q) > -\infty$; otherwise it is $0$. Similarly for  
$\La \ln V(\D) \Ra_{\D}$, using that this is the derivative w.r.t to
$n$ of  $\ln \La V^n(\D) \Ra_\D$ at $n=0$, one has
\begin{equation*}
\lim_{N\rightarrow\infty}N^{-1}\La \ln V(\D) \Ra_{\D} = \min_q\, f_1(0,q)\,.
\end{equation*}

To find the critical value $\alpha_c$ where the minimum of $f_1(0,q)$ 
diverges to $-\infty$, note that $f_1(0,q)$ only diverges for 
$q=1$.
The average
$\La \ln H(-z_0 \sqrt q/\sqrt{1-q}\,)  \Ra_{z_0}$ can be calculated 
analytically for $q\rightarrow 1$ using that for large positive arguments 
$H(x) \sim \frac{\exp(-\half x^2)}{x\sqrt{2 \pi}}$, whereas 
$H(-\infty) =1$. This yields that for $q\rightarrow 1$
\begin{equation*}
\La \ln H(-z_0 \sqrt q/\sqrt{1-q}\,)  \Ra_{z_0} \sim - \frac{1}{4}\frac{q}{1-q} 
\end{equation*}
and the final result that $\min_q\, f_1(0,q)$ is only finite if 
$\alpha < \alpha_c = 2$.

In the limit of large $N$,  almost all dichotomies 
can be implemented by the perceptron up to $\alpha_c$, 
but the fraction of implementable 
dichotomies is vanishingly small when $\alpha > \alpha_c$. In this
sense Gardner's calculation exactly coincides with the result obtained
by Schl\"afli. Further the fact that $\La \Theta(V(\D)) \Ra_{\D}$ converges
to a step function, implies that $\Theta(V(\D))$ is selfaveraging:
for any $\epsilon > 0$ the probability that 
$|\Theta(V(\D)- \La \Theta(V(\D)) \Ra_{\D}| > \epsilon$ vanishes in the
large $N$ limit (except at $\alpha_c =2$). A simple scaling argument shows
that $N^{-1}\ln V(\D)$ should be selfaveraging as well. The variance
of $\ln  V(\D)$ is equal to the second derivative w.r.t. to $n$ of 
$\ln\La V^n(\D)\Ra_{\D}$ evaluated at $n=0$.  When using replicas to 
find  an analytical   continuation to small $n$, one must obtain
that the second derivative of $N^{-1}\ln\La(V^n(\D)) \Ra_{\D}$ is finite.
This yields that the variance of $N^{-1}\ln V(\D)$ is ${\cal O}(1/N)$.

To round off Gardner's calculation we consider the physical interpretation
of the parameter $q$, relating it to the mean of all weight vectors 
implementing a given dichotomy:
\begin{equation*}
\bar{J}(\D) = V^{-1}(\D)
               \int \!{\rm d}J \,J\, \prod_{\mu=1}^{\alpha N}
                      \Theta(\tau^\mu J^T\xi^\mu).
\end{equation*} 
One can show that $\La \| \bar{J}(\D) \|^2 \Ra_D \rightarrow q^*$ with
increasing $N$, where $q^*$ is the value minimizing $f_1(0,q)$.
Further, the averaged squared length of $\bar{J}(\D)$ can be written as
\begin{equation*}
\La \| \bar{J}(\D) \|^2 \Ra_D 
= \La  V^{-2}(\D)
     \int \!{\rm d}J^1\!{\rm d}J^2\, {J^1}^T\!\!J^2\, 
     \prod_{\mu=1}^P \prod_{a=1}^2
     \Theta(\tau^\mu {J^a}^T\xi^\mu) \Ra_D \,.
\end{equation*}
It can thus also be regarded as the average overlap of two perceptron
weight vectors $J^1$ and $J^2$, picked at random from
the set of all perceptrons which implement a given dichotomy. This physical
interpretation of $q$ will be derived in Section 5.3, in the context
of discussing the relationship between the parameterization of the matrix $Q$
and the distribution of the overlaps ${J^1}^T\!\!J^2$. As a consequence we 
shall also find that $\| \bar{J}(\D) \|^2$ is selfaveraging if the replica
symmetric parameterization is correct.

%% file: extram.tex
\chapter{Extensions and Ramifications}

\section{Beyond capacity}

If the number of patterns in the training set $\D$ is too large, no perceptron
will exist which implements the dichotomy perfectly. One may, however, still
try to find a network $\sigma$ which makes few mistakes on $\D$, 
so $\sigma$ should have a small training error
\begin{equation}
\epsilon_\D(\sigma) = 
  P^{-1}\sum_{\mu=1}^P \Theta(-\tau^\mu\sigma(\xi^\mu))\,.
\end{equation}
To adapt Gardner's calculation one first defines a probability 
density on the class of all networks by
\begin{equation}
p(\sigma) = \frac{e^{- P\beta \epsilon_D(\sigma)}}{Z(\D)}\,, \label{Temp}
\end{equation}
where the partition function $Z(\D)$ assures that the density is normalized.
The parameter $\beta$ is called the inverse temperature,  
and a network drawn from 
$p(\sigma)$ will have minimal training error in the limit of
large $\beta$. Note that one can calculate the average, w.r.t. $p(\sigma)$,
of the training error from the derivative of  
$\ln Z(\D)$ with respect to $\beta$.

In the context of a dynamical interpretation, $p(\sigma)$ is 
the stationary distribution of a suitable Langevin dynamics and 
$-\beta^{-1}\ln Z(\D)$ plays the r\^ole of a free energy. But I shall
not be concerned with such an interpretation here. 

Choosing $F$ as
\begin{equation*}
F(x) = e^{-\beta  \Theta(-x)}\;,
\end{equation*}
in the case of the perceptron yields
\begin{equation}
 Z(\D) = \int\!{\rm d}J e^{-\beta P \epsilon_D(\sigma_J)}  =  
\int \!{\rm d}J \prod_{\mu=1}^P F(\tau^\mu J^T\xi^\mu)\,.
\label{part2}
\end{equation}
The second expression is the same as in the definition of $Z(\D)$
used in Gardner's calculation, Eq. (\ref{part}). So, in replica symmetry,
we have already calculated $N^{-1}\La \ln Z(\D) \Ra_{\D}$.
However, major complications arise from the fact that the assumption
of replica symmetry is wrong for $\alpha > \alpha_c$. This is extensively
reviewed in \cite{Gyo01}. I shall discuss the techniques for dealing with
broken replica symmetry in the context of multilayer networks.

Considering the case of finite $\beta$ helps to deal with a technical
problem in the above exposition of Gardner's calculation. 
From Schl\"aflis result we know that for finite $N$ and $\alpha > 1$
there is a finite probability that $V(\D) =0$. So 
$\La \ln V(\D) \Ra_D$ diverges, and the expression we found for the large
$N$ limit of $N^{-1}\La \ln V(\D) \Ra_D$ is surely {\em incorrect} in the 
range 
$1 < \alpha < 2$. This is actually quite pleasing since our result reflects
the fact that the probability that  $V(\D)=0$ vanishes in the large $N$
limit for $\alpha < 2$. To make sense of the calculations, however, one
should use Eq. (\ref{part2}) at a finite value of $\beta$ and when
considering $N^{-1}\La \ln Z(\D) \Ra_D$ first take the limit of large $N$
and then the limit $\beta\rightarrow\infty$. In the replica calculation
the two limits commute since for finite $N$ we are only calculating
the moments of $Z(\D)$.

{

\newcommand{\LL}{{\mathbb L}}
\section{Discrete weight vectors}

Upto now we have assumed that the components of the weight vector
can take on any real value. In numerical calculations, however, the set
of possible values will be finite, even if it can be quite large.
We thus assume that the vector $J$ is restricted to
lie in a finite subset $\LL$ of $\R^N$ and consider the
number $M_\LL(\D)$ of networks from $\LL$ that implement a given
dichotomy:
\begin{equation*}
M_\LL(\D) = \sum_{J\in\LL} \prod_{\mu=1}^P \Theta(\tau^\mu J^T\xi^\mu)\,.
\end{equation*}

As for continuous weights it is again instructive and simple to calculate the
average,
$\La M_\LL(\D) \Ra_\D = 2^{-P} {\rm card}\, \LL$. 
For $P = \alpha N$ the average will become zero for large $N$ unless the 
number of networks increases at least exponentially with $N$.  
We thus assume that
there are $L$ possible values for each weight and so ${\rm card}\, 
\LL = L^N$.
Then for large $N$ one finds  
$\La M_\LL(\D) \Ra_\D = 0$ when $\alpha > \log_2 L$. Since the possible
values of  $M_\LL(\D)$ are discrete, for such an $\alpha$
the probability that a dichotomy can be implemented by a network in $\LL$
becomes zero. This is sometimes called the information theoretic bound,
since any weight vector in $\LL$ can be represented using $N \log_2 L$ bits.  

It is interesting that in contrast to continuous weights
a simple average of $M_\LL(\D)$ can yield some insight into the critical 
capacity. But already for $L > 4$ a tighter bound for the perceptron
is $\alpha_c \leq 2$, treating the discrete case as a restriction of the
continuous one. To improve on the information theoretic bound, let $\phi$
be an orthogonal transformation of $\R^N$ and $\phi\LL$ the set 
obtained applying $\phi$ to the elements of $\LL$. Since the distribution
of inputs is isotropic, for any function $f$:
\begin{equation*}
\La f(M_\LL(\D)) \Ra_\D = \La f(M_{\phi\LL}(\D)) \Ra_\D\;.
\end{equation*}
Denote by $\La \ldots  \Ra_\phi$ the average over the uniform density on the 
orthogonal group of $\R^N$, then for any convex function $f$:
\begin{equation*}
\La f(M_\LL(\D)) \Ra_\D = \La\La f(M_{\phi\LL}(\D)) \Ra_\D\Ra_\phi
\leq \La f(\La M_{\phi\LL}(\D)\Ra_\phi) \Ra_\D
\end{equation*}
Now $\La M_{\phi\LL}(\D) \Ra_\phi = V(\D){\rm card}\, \LL$ and we obtain
a simple bound in terms of the spherical volume. In particular
if $\LL$ has $L^N$ elements we have 
$\La M^n_\LL(\D)) \Ra_\D  \leq L^{Nn} \La V^n(\D)) \Ra_\D$
for $0 < n \leq 1$. 
Since $M_\LL(\D)$ is integer, 
$\Theta(M_\LL(\D)) \leq  M^n_\LL(\D)$, and we obtain an upper bound
on the probability that a dichotomy can be implemented by one of the $L^N$
vectors 
\begin{equation*}
\La \Theta(M_\LL(\D)) \Ra_\D \leq L^{Nn} \La V^n(\D)) \Ra_\D\;.
\end{equation*}
For $n=1$ this is just the information theoretic bound but tighter
bounds can be obtained by evaluating the RHS for $n<1$.
Using the results for the continuous case, one can easily compute the smallest
$\alpha_c(L)$ with the property 
that for any $\alpha > \alpha_c(L)$ the right hand side
decays to zero exponentially with increasing $N$ for some {\em finite} value
$n(\alpha)$. This yields an upper bound on the critical capacity.
For $L=2$ one obtains $\alpha_c(2) = 0.85$ and this bound is close
to the value $\alpha_c = 0.83$ found for $\LL = \{-1,1\}^N$ by 
calculating $N^{-1}\La \ln M_\LL(\D)) \Ra_\D$ using replicas 
\cite{Kra89}. The latter value is in good agreement with results from
numerical simulations (\schrseven). Further, based on the findings in
\cite{urb94} for equidistant weight values, one will expect the bound 
$\alpha_c(L)$
to be asymptotically tight for large $L$.

From a conceptual point of view the case of discrete weight is nice because
an assumption implicit in the interpretation of Gardner's calculation
can be avoided. In identifying the critical capacity with the divergence
of $N^{-1}\La \ln Z(D) \Ra_D$ as well as  in commuting the limit 
$n\rightarrow 0$  with $N\rightarrow\infty$ in the calculation of
$\La \Theta(V(D)) \Ra_D$ , one assumes that for an implementable dichotomy 
$\ln V(\D)$ typically is on the order of $N$.  In the discrete case this 
assumption is not needed. The reason for this is of course that the 
information theoretic bound guarantees that below the capacity limit
$\ln M_\LL(\D))$ is on the order of $N$.

}

\section{More general input distributions}

Up to now we have assumed that the components of $\xi$ are independent
and Gaussian. But in Gardner's calculation the essential point is not that
$\xi$ is Gaussian but that the field $J^T\xi$ is Gaussian. Using the central
limit theorem, one can argue that this will also  be the case when the input 
components are not Gaussian but just independent. If further the components 
have zero mean and unit variance, Gardner's calculation does not even have to 
be modified.

It is, however, worthwhile noting that the central limit theorem will not
apply for all choices of $J$. While the set of exceptions will have zero 
measure for large $N$, so does the version space, the set of weight 
vectors implementing a given dichotomy. Reasonably, one will not expect
this to be a problem; but it would be difficult to actually show that
all the important contributions to $\La \ln Z(\D) \Ra_D$ do come from the 
region of state space where the central limit theorem holds. 

If one is prepared to live with this, one can argue that the 
assumption of independent input components is too strong.%
\footnote{ The following argument is due to Manfred Opper, personal
           communication.}
We consider the characteristic function of $J^T\xi$
\[
c_J(k) = \La e^{{\rm i}J^T\xi} \Ra_\xi\, 
\]
and for simplicity assume that $J$ is drawn from a isotropic Gaussian
distribution with the normalization $\La \|J\|^2 \Ra  = 1$. Then, for the 
average value of $c_J(k)$ one immediately finds
\[
\La c_J(k) \Ra_J = \La e^{-\half k^2 \|\xi\|^2/N} \Ra_\xi\,.
\] 
We now assume that $\|\xi\|^2$ is selfaveraging with mean $N$ and not
too malicious, so that $\La c_J(k) \Ra_J = e^{-\half k^2}$, 
i.e. a Gaussian. Generically $c_J(k)$ will thus also be Gaussian if
$c_J(k)$ is selfaveraging. For its second moment we obtain
\[
\La c_J(k)^2 \Ra_J = \La e^{-\half k^2 \|\xi^1+\xi^2\|^2/N} \Ra_{\xi^1,\xi^2}
         = \La e^{-\half k^2  (\|\xi^1\|^2+ \|\xi^2\|^2)/N}
               e^{-k^2 {\xi^1}^T\xi^2/N}
            \Ra_{\xi^1,\xi^2}\,,
\]
where $\xi^1$ and $\xi^2$ are independent and have the same distribution
as $\xi$. For a large class of distributions, ${\xi^1}^T\xi^2$ is sufficiently
small compared to $N$ so that for large N:
\[
\La c_J(k)^2 \Ra_J \rightarrow 
\La e^{-\half k^2  (\|\xi^1\|^2+ \|\xi^2\|^2)/N} \Ra_{\xi^1,\xi^2} =
\La c_J(k)\Ra_J^2\;.
\]
In this case, the variance of $c_J(k)$ vanishes for large $N$,
$c_J(k)$ is selfaveraging, and typically $J^T\xi$ becomes Gaussian.

Even if one will thus expect that it is not really necessary,
in the sequel I shall nevertheless assume i.i.d.  ${\cal N}(0,1)$ input 
components for brevity and simplicity.

\chapter{Learning a rule }

In the capacity problem one assumes that the outputs in the training set
are independent of the inputs. For pattern recognition, however, one is 
mainly interested in the performance of the network on inputs which were 
not used for training. 
This only makes sense if one assumes that the desired output
is not random but depends on the input, and that this dependency is learned
by the network based on the training examples. 
So while the training data still consists of $P$ independent samples of the
random variable $(\xi,\tau)$, one no longer assumes that $\tau$ is independent
of $\xi$.
One scenario is that 
the desired output $\tau$ is a binary function $b(\xi)$ of the input $\xi$, and 
$b$ is then sometimes called the teacher.
One can then measure how well a student, i.e.  a network $\sigma$,
approximates the input/output relationship by defining 
the generalization error
\begin{equation}
\epsilon_g(\sigma) = \La \Theta(-\tau\sigma(\xi)) \Ra_{(\xi,\tau)}\,,
\end{equation}
which is just the probability that $(\xi,\sigma(\xi)) \neq (\xi,\tau)$. 
Training then amounts to finding a student which makes 
few mistakes on the examples $\D$, and this is measured by the training 
error
\begin{equation*}
\epsilon_\D(\sigma) = 
  P^{-1}\sum_{\mu=1}^P \Theta(-\tau^\mu\sigma(\xi^\mu))\,.
\end{equation*}
A key question in formal learning theory is to which extent
minimizing $\epsilon_\D$ is conducive to the actual goal of
minimizing $\epsilon_g$.

In the case that the network is a perceptron, as simple model is
that $b$ can be implemented by a perceptron, i.e.
\begin{equation*}
b(\xi) = \sign(B^T\xi)
\end{equation*}
with  a suitable weight vector $B\in\R^N$. If the input components
are i.i.d. $N(0,1)$ as in the capacity problem, it is simple to
calculate the generalization error of a perceptron $\sigma_J$
\begin{equation}
\epsilon_g(\sigma_J) = \La \Theta(-B^T\xi J^T\xi)) \Ra_\xi\ =
\frac{1}{\pi}\arccos B^TJ\,.
\label{Peg}
\end{equation}
Here, and in the sequel, the weight vectors are normalized to $1$.

Since the teacher is a perceptron, in contrast to the capacity problem,
$V(\D)$ is nonzero for all training set sizes $P$. One can, however,
ask whether a student $J$ can generalize badly but still achieve zero 
training error.  In view of Eq. (\ref{Peg}) generalizing badly means
that $B^TJ$ is small and one thus consider the restricted volumes
\begin{equation*}
V_R(\D) = \int \!{\rm d}J \,\delta(R - B^TJ) 
\prod_{\mu=1}^P \Theta(\tau^\mu J^T\xi^\mu)\,.
\end{equation*}
When $\alpha=P/N$ is sufficiently large, one will expect that 
$\Theta(V_R(\D))$ vanishes  unless $R$ exceeds a critical
value $R_c$. So a student 
with zero training error must have  a generalization error smaller 
than $\frac{1}{\pi}\arccos R_c$ and this is sometimes called
the worst case generalization behavior. One may also consider
the expected generalization behavior by computing 
the value $R^*$ which maximizes $\ln V_R(\D)$.
This yields the most probable generalization error of a student picked
at random among all perceptrons with zero training error. Since, as in the
capacity problem, $N^{-1}\ln V_R(\D) $ is on the order of $1$ unless
it diverges, for large $N$ the volume corresponding to $R^*$ is much 
larger than that of any other value of $R$, and the most probable 
generalization error is in fact observed with probability $1$ in this limit.

By the same arguments as in the capacity problem $\Theta(V_R(\D))$ and
$N^{-1}\ln V_R(\D)$ are selfaveraging, and the generalization behavior
is obtained by a straightforward adaptation
of Gardner's calculation. We again consider the more general form
\begin{equation}
Z_R(\D)  = 
\int \!{\rm d}J \,\delta(R - B^TJ) \prod_{\mu=1}^P F(\tau^\mu J^T\xi^\mu)\,
\label{Rpart}
\end{equation}
and obtain for its moments
\begin{equation}
\La Z_R^n(\D) \Ra_\D  = \int \!\dd{\J} \La  \prod_{a=1}^n  
                       F(B^T\xi {J^a}^T\xi) \Ra_{\!\xi}^P 
                       \prod_{a=1}^n \delta(R - B^TJ^a) \,.
\label{tmom}
\end{equation} 
The $n+1$ random variables $X_{n+1} = B^T\xi$ and $X_a = {J^a}^T\xi$ are 
Gaussian with a covariance matrix
\begin{equation*}
\hat{Q} = \left( \begin{array}{cc} Q & \bf R \\ \bf{R}^T & 1 \end{array} \right )
\,,
\end{equation*}
here $Q$ is the $n$ by $n$ matrix  $Q_{ab} ={J^a}^TJ^b$ and 
${\bf R} = (R,R,\ldots,R)^T$ is an $n$-dimensional vector.  So
\begin{equation*}
\La Z_R^n(\D) \Ra_\D  = \int \!\dd{Q}  D_n(Q,R)
\La  \prod_{a=1}^n F(X_{n+1}(\hat{Q}) X_a(\hat{Q}) ) \Ra_{X(\hat{Q})}^P \,,
\end{equation*}
where 
$D_n(Q,R) = \int \!\dd{\J}\, \delta( Q - \J^T \J)
                             \prod_{a=1}^n \delta(R-B^TJ^a)$.

To calculate $D_n(Q,R)$, note that by rotational symmetry
it is invariant to the choice of $B$
as long as $\|B\|=1$. So averaging over the uniform density on the unit
sphere yields
\begin{equation}
D_n(Q,R) = \int {\rm d}B\, D_n(Q,R) = D_{n+1}(\hat{Q}) = 
D_{n+1}({\bf 1}) (\det\hat{Q})^{(N-n-2)/2}\,.
\end{equation}
The expression for $\det\hat{Q}$ can be simplified since
for a square block matrix 
$(\begin{smallmatrix} a & b \\ c & d \end{smallmatrix})$ with invertible
square matrices $a$ and $d$
\begin{equation}
\det\left(\begin{smallmatrix} a & b \\ c & d \end{smallmatrix}\right)  = 
\det(a - b\, d^{-1} c) \det d
\label{deteq1}
\end{equation}
holds
\footnote{
Note that 
$
\left(\begin{smallmatrix} \bf 1 & -b \\ \bf 0  & \bf 1 
\end{smallmatrix}\right)
\left(\begin{smallmatrix} \bf 1 & b \\ \bf c  & \bf 1 
\end{smallmatrix}\right)
=
 \left(\begin{smallmatrix} {\bf 1}-bc & \bf 0 \\ \bf c  & \bf 1 
\end{smallmatrix}\right)
$.
In this equation, it is trivial to obtain the determinant of two
matrices, and thus the third. Hence, the same statement applies to
$
\left(\begin{smallmatrix} a^{-1} & \bf 0 \\ \bf 0  & d^{-1} 
\end{smallmatrix}\right)
\left(\begin{smallmatrix} a & b \\ \bf c  & d 
\end{smallmatrix}\right)
=
 \left(\begin{smallmatrix} {\bf 1} &  a^{-1}b \\ d^{-1}c  & \bf 1 
\end{smallmatrix}\right)\,.
$
}.
Thus $\det\hat{Q} = \det( Q - \bf R {\bf R}^T)$.

To evaluate Eq. (\ref{tmom}) for large $N$ and $P=\alpha N$, we again
assume that the value of $Q$ maximizing the integrand is replica symmetric,
i.e. $Q = M_n(1,q)$. So it is straightforward
to evaluate $D_n(Q,R)$ since $\det\hat{Q} = \det M_n(1-R^2,q-R^2)$.
Further, at the maximum
$X(Q)$ can be rewritten in terms of i.i.d. $N(0,1)$ random variables
$z_{-1},z_0,\ldots,z_n$ as
\begin{equation*}
X_a(\hat{Q}) =  R z_{-1} + \sqrt{q-R^2}z_0 + \sqrt{1-q}z_a \quad\mbox{and}\quad
X_{n+1}(\hat{Q}) = z_{-1}\,.
\end{equation*}
Then at the maximum the average in Eq. (\ref{tmom}) simplifies to
\begin{eqnarray*}
\lefteqn{
\La  \prod_{a=1}^n F(X_{n+1}(\hat{Q}) X_a(\hat{Q}) ) \Ra_{X(\hat{Q})}
= } \nonumber \\
& &
\La\La F\left(z_{-1}(R z_{-1} + \sqrt{q-R^2}z_0 + \sqrt{1-q}z_1)\right)
\Ra_{z_1}^n \Ra_{z_{-1},z_0},
\end{eqnarray*}
where the second expression makes sense also for noninteger $n$.

As in the capacity problem one now uses an analytical continuation
to find for small $n$
\begin{equation}
\lim_{N\rightarrow\infty}N^{-1} \ln \La Z_R^n(\D) \Ra_\D  = 
{\cal O}(n^2) + n \min_q\, g(R,q)\,,
\label{Tsmom}
\end{equation}
with
\begin{eqnarray*}
g(R,q) &=& \half \left( \frac{q-R^2}{1-q} + \ln(1-q) \right) + \nonumber \\
&&\alpha\La \ln \La 
F\left(z_{-1}(R z_{-1} + \sqrt{q-R^2}z_0 + \sqrt{1-q}z_1)\right)
\Ra_{z_1} \Ra_{z_{-1},z_0} .
\end{eqnarray*}

Specializing to the case that $F(x) = \Theta(x)$ allows the minor
simplification of rewriting the $z_1$-average in terms of $H(x)$.
Then the analysis of (\ref{Tsmom}) shows that the generalization error
decays to zero as $1/\alpha$ in the worst case as well as in the
expected case and that only the pre-factor differs in the two cases
\cite{Eng93,Gyo90}.
However, a much larger difference between the two scenarios has been
found for some multilayer networks (\urbsix).

That the teacher is a perceptron, is a rather unrealistic assumption.
In a more general case it may be impossible to find a network
which has zero training error. A reasonable strategy is then to look
for a network with minimal training error. 

A simple model of such a situation is that student and teacher are 
perceptrons but the output of the teacher is corrupted by noise. The cases of
additive and  multiplicative noise have been widely analyzed 
\cite{Gyo90,Opp91,Opp91b,Wat93}. In the
first case $\tau = \sign(B^T\xi + \eta)$ where the noise term $\eta$
is independent of $\xi$ and typically assumed $N(0,\nu)$. For multiplicative
noise $\tau = \sign(B^T\xi)\eta$, $\eta$ is $\pm 1$ and again independent
of  $\xi$. In this case one will reasonably assume that the mean of $\eta$ is 
positive so that $\tau$ equals the uncorrupted output $\sign(B^T\xi)$  
with a probability greater than $1/2$.

It is easy to apply the above analysis to the noisy cases since
the generalization error of a perceptron $\sigma_J$
is still just a function of the overlap $R = B^T J$. 
To consider the Gibbs density $p(\sigma)$ given by Eq. (\ref{Temp}))
we use $F(x) = e^{-\beta P \Theta(-x)}$  in the definition of $Z_R(\D)$.
The partition
function then is $Z(\D) = \int_{0}^1 {\rm d}R\, Z_R(\D)$. Now the probability
that the weight vector of a student $\sigma_J$ drawn from $p$
has an overlap $R=B^TJ$ with the teacher is $Z_R(\D)/Z(\D)$,  and the most
probable value $R^*$ of $R$ is obtained by maximizing $Z_R(\D)$. 
Since $N^{-1} \ln Z_R(\D)$ and hence $N^{-1} \ln Z(D)$ are selfaveraging,
in the thermodynamic limit $R^*$ is again obtained by maximizing 
$\La N^{-1}\ln Z_R(\D) \Ra_D$.

Major complications arise from the fact that the replica symmetric
assumption is invalid for sufficiently high $\beta$ if no student with
zero training error exist \cite{Gyo90}. This problem can probably be avoided
within the framework of a Bayesian analysis, which yields that in the 
presence of noise 
the training error should not be minimized when aiming for good 
generalization. 
Instead, in the case of multiplicative noise, one uses a 
carefully chosen finite value of the inverse temperature. Unfortunately
this value depends not only on $\alpha$ but also on 
the noise level \cite{Opp91,Opp91b}. Since the Bayesian strategy involves
many assumptions about what is being learned, one may wish to stick with
the suboptimal but generally applicable strategy of minimizing $\epsilon_\D$,
and I shall shortly describe techniques for dealing with the broken replica
symmetry.

%% file: mlp.tex
\chapter{Multilayer perceptrons}

As mentioned in the introduction a general two layer network is given
by
\begin{equation*}
\sigma_{J,w}(\xi) = g\left( \sum_{k=1}^K w_k h(\xi^T J_k)\right)\,,
\end{equation*}
and such networks have found many applications both in regression 
and classification problems. 
In statistical physics it has only been possible to analyze these networks
in the limit where the number of input dimensions $N$ is much larger
than the number of hidden units $K$. In this limit one will not expect
the few adaptable hidden to output couplings $w_k$ to play a major
r\^ole. Hence one considers so called committee machines where the
$w_k$ are constant and equal $1$. (Sometimes, for the sake of normalization,
one assumes $w_k = 1/\sqrt{K}$ instead.)

Formally, the analysis of regression and classification is very similar, and
for brevity I shall consider only classification here, 
results for regression can be found in the papers 7 and 8. 
For classification the output function is $g(x) = \sign(x)$ and I shall
also assume that $h$ is the sign function. So in the sequel the term committee
machine (CM) refers to the class of functions
\begin{equation}
\sigma_{J}(\xi) = \sign\left( \sum_{k=1}^K \sign(\xi^T J_k)\right)\,.
\label{com}
\end{equation} 
Note that, as for a real committee, the output is decided by the majority
vote of the $K$ hidden units, and we shall assume that $K$ is odd 
to avoid a draw.
Sometimes it is convenient to consider
a simplified architecture the so called tree committee machine (TCM). 
For the tree
the input $\xi$ is $NK$ dimensional, composed of $K$ vectors
$\xi_k \in \R^N$, and
\begin{equation}
\sigma_{J}(\xi) = \sign\left( \sum_{k=1}^K \sign(\xi_k^T J_k)\right)\,.
\end{equation} 
This is simpler because the fields $\xi_k^T J_k$ are now statistically 
independent if all input components are independent.%
\footnote{ In much of the literature the definition of the tree committee 
assumes $N/K$ dimensional $J_k$ and $\xi_k$, so that the number of free 
parameters is $N$ and not $NK$ as in the above definition. But this difference
is immaterial as long as final results are expressed in terms of the 
ratio of examples to free parameters.} 

For the committee machine (CM) it is interesting to consider the effect
of correlations between the fields $\xi^T J_k$. Let us assume that 
$J_k = p w_0 + \sqrt{1-p^2} w_k$ with orthonormal vectors $w_j$ and $p >0$. 
Then one will expect that quite often $\sigma_{J}(\xi) = \sign(\xi^T w_0)$.
Indeed, by the law of large numbers, $\sigma_{J}(\xi) = \sign(\xi^T w_0)$,
will hold with a probability approaching $1$ in the limit of large $K$
if the input components are i.i.d. $N(0,1)$. So for any finite value of the
correlation $p$ the output of the committee becomes identical to that of
a perceptron with weight vector $w_0$ in this limit. Hence in many
contexts one will expect $p$ to be small when $K$ is large.

We now turn to the capacity problem for these architectures setting
\begin{equation}
Z(\D)  = \int \!{\rm d}J \prod_{\mu=1}^P F(\tau^\mu \sigma_J(\xi^\mu))\,,
\label{mpart}
\end{equation}
and as for the perceptron
\begin{equation*}
\La Z^n(\D) \Ra_\D  = \int \!\dd{\J} \La  \prod_{a=1}^n  
                       F(\tau \sigma_{J^a}(\xi)) \Ra_{(\xi,\tau)}^P \,.
\end{equation*}
Now $\dd{\J}$ refers to an integration over $Kn$ unit spheres in $\R^N$.
Further we define the $Kn$ by $Kn$ order parameter matrix ${\bf Q}$ as
${\bf Q}^{ab}_{kl} = {J^a_k}^T J^b_l$ 
for $a,b = 1,\ldots,n$ and $k,l = 1,\ldots,K$ and an $Kn$ dimensional 
Gaussian $X({\bf Q})$ with zero mean and covariances
\begin{equation*}
\La X^a_k({\bf Q}) X^b_l({\bf Q}) \Ra  = 
\left\{\begin{array}{rl}             {\bf Q}^{ab}_{kl} &\quad  \mbox{CM} \\
                       \delta_{kl} {\bf Q}^{ab}_{kl} &\quad \mbox{TCM}\,.
\end{array}\right. 
\end{equation*}
The value of $\sigma_{J^a}(\xi)$ is determined by the  values of
$\xi^TJ^a_k$ for the CM and by $\xi_k^TJ^a_k$ for the TCM. Thus
\begin{equation*}
\La  \prod_{a=1}^n F(\tau \sigma_{J^a}(\xi)) \Ra_{(\xi,\tau)} = 
\La  \prod_{a=1}^n F\left(\sign(\sum_{k=1}^K \sign(X^a_k({\bf Q}))) \right)
\Ra_{X({\bf Q})}\!\!.
\end{equation*}
Defining the load parameter as $\alpha = \frac{P}{KN}$ and using the same
arguments as in the case of the perceptron yields
\begin{equation}
\lim_{N\rightarrow\infty}\frac{\ln \La Z^n(\D) \Ra_\D}{KN}  = 
\max_{\bf Q}\, \frac{1}{2K}\ln\det{\bf Q}  +
\alpha \ln \La \prod_{a=1}^n F\left(\sign(\sum_{k=1}^K \sign(X^a_k({\bf Q}))) 
           \right) \Ra_{X({\bf Q})}\!\!.
\label{mrep}
\end{equation}

In principle one could now adopt an, e.g.,  replica symmetric parameterization
of ${\bf Q}$ and obtain an analytic continuation to small $n$.
One then still  has a large number of order parameters and the extremal
problem involves a $K$-fold integral which has to be done numerically. 
While it would probably be feasible to solve this problem for $K=3$,
to my knowledge no one has done this.   

To simplify the extremal problem, it is convenient to view  
${\bf Q}$ as a $K$ by $K$ block matrix indexed by the site indeces
$k,l$ and consisting of blocks which are $n$ by $n$ matrices.
For the TCM the energy term does not depend on ${\bf Q}^{ab}_{kl}$
if $k\neq l$ and thus in this case at the maximum 
${\bf Q}^{ab}_{kl} = 0$.%
\footnote{ 
It suffices to show this for a symmetric $2$ by $2$ block matrix
 $U=(\begin{smallmatrix} a & c \\ c^T & b \end{smallmatrix})$, the extension
to general $K$ is then by induction. Using (\ref{deteq1}) one has 
$\det U = \det (\begin{smallmatrix} a & -c \\ -c^T & b \end{smallmatrix})$.
So over the positive definite matrices,
$0 = \arg\max_c\det(\begin{smallmatrix} a & c \\ c^T & b \end{smallmatrix})$, 
since $\ln \det U$ is a convex function on these matrices.
The convexity can be shown by rewriting 
$(\det (\lambda U + (1-\lambda) V))^{-1/2}$
as a Gaussian integral and applying H\"olders inequality. 
}
For both architectures we now make the site symmetric assumption that at the
extremum:
\begin{equation}
{\bf Q}^{ab}_{kl} =\delta_{kl} Q^{ab} + P^{ab}/K, \label{sitesym}
\end{equation}
which can be more concisely written in block form: 
${\bf Q} = M_K(Q+P/K,P/K)$. As just noted $P=0$ for the TCM. 

Now $\det {\bf Q}$  can be evaluated in a way which is analogous to the 
calculation of $M_n(u,v)$. Let $U,V$ be n by n matrices and $x\in\R^N$ then
\begin{equation*}
M_K(U,V) \left( \begin{array}{c} x \\ x\\ \vdots \\ x  \end{array} \right ) =
\left.
\left( \begin{array}{c}(U+(K-1)V) x \\ (U+(K-1)V)x\\ \vdots \\(U+(K-1)V) x  \end{array} 
\right )\quad\right\} K  \mbox{ rows} 
\end{equation*}
as well as
\begin{equation*}
M_K(U,V) \left( \begin{array}{r} x \\ -x\\ 0\\ \vdots \\ 0  \end{array} 
\right ) =
\left( \begin{array}{c}\hfill(U-V) x \\-(U-V)x\\ 0\\ \vdots \\ 0  \end{array} 
\right)\,.
\end{equation*}
The last equation stays valid if the rows of the argument vector and the
resulting vector are permuted.
We thus obtain a decomposition of $\R^{Kn}$ into a direct sum of $K$ 
$n$-dimensional
eigenspaces of $M_K(U,V)$  and the determinant of $M_K(U,V)$  
is just the product of the determinants on the eigenspaces:
\begin{equation}
\det M_K(U,V) = \det(U+(K-1)V) \det(U-V)^{K-1}\;, \label{decomp}
\end{equation}
and in particular 
\begin{equation}
\det {\bf Q} = \det(Q+P)\det Q^{K-1}. \label{mlpent}
\end{equation}

Equations (\ref{mrep}, \ref{mlpent}) form the basis for the following 
discussion
of committee machines. I shall first consider a limiting scenario in which
for large $K$ the summation over hidden units is exploited to simplify the
energy term using the central limit theorem. This leads to a very simple
Gaussian theory which is formally similar to the one for the perceptron. The
main result is that the storage capacity diverges with $K$, but the theory
only yields limited insight into the rate of divergence. But this approach is 
highly suited to the analysis of learning problems where the target outputs 
are not random but given by a rule 
(see \cite{Schw92,Schw92a,Schw93a,Schw93b,Urb95,Urb95b} and \urbsevenb). 
As in the case
of the perceptron, adapting the capacity calculations to a learning problem
is relatively straightforward and I shall not dwell on this. Instead, in the
next chapter,  I shall consider the more precise capacity calculation 
obtained by taking the $n\rightarrow 0$ limit for fixed $K$ and the 
interpretation of the replica symmetry breaking found in this calculation in 
terms of the internal representations of the committee machine. 

\section{Gaussian theory of committee machines }

The main idea here is to simplify  the energy term by arguing
that the distribution of
\begin{equation*}
Y_a = K^{-1/2} \sum_{k=1}^K \sign(X^a_k({\bf Q}))
\end{equation*}
becomes Gaussian for large $K$. For the TCM this is just stating the  
multidimensional central limit theorem since  $X^a_k({\bf Q})$ and 
$X^b_l({\bf Q})$ are independent if $k \neq l$. This does not hold for the CM,
but assuming the site symmetric parameterization it has been shown in
\cite{Urb95b} that the limiting joint distribution of the $Y_a$
is Gaussian. Nevertheless, to reduce clutter, I shall only consider
the TCM in this section.

Obviously the mean of $Y_a$ is zero and for the covariances one
has
\begin{equation*}
\La Y_a Y_b \Ra =
K^{-1}  \sum_{k=1}^K \La \sign(X^a_k({\bf Q}) \sign(X^b_k({\bf Q}) \Ra =
K^{-1}  \sum_{k=1}^K \frac{2}{\pi} \arcsin({\bf Q}^{ab}_{kk}) =
 \frac{2}{\pi} \arcsin(Q^{ab})\;.
\end{equation*}
I have assumed site symmetry for the last equality. So from 
(\ref{mrep},\ref{sitesym},\ref{mlpent}) we obtain
\begin{equation}
\lim_{K,N\rightarrow\infty}\frac{\ln \La Z^n(\D) \Ra_\D}{KN} =
\max_Q\, \half\ln\det Q +
\alpha \ln \La  \prod_{a=1}^n F(\sign( Y_a(Q^{\rm e})) ) \Ra_{Y(Q^{\rm e})},
\label{TCMn}
\end{equation}
where $Y(Q^{\rm e})$ is an $n$-dimensional Gaussian with zero mean and
\begin{equation*}
\La Y_a(Q^{\rm e}) Y_b(Q^{\rm e}) \Ra = (Q^{\rm e})^{ab} = 
\frac{2}{\pi} \arcsin(Q^{ab}).
\end{equation*}
The essential difference to the corresponding expression for the
perceptron (\ref{Glab}) is that in the energy term the correlation
matrix $Q$ is replaced by the effective correlations $Q^{\rm e}$.
   
Adopting a replica symmetric parameterization of $Q$, it is now straightforward
to take the limit of small $n$ and for $F=\Theta$ one finds
\begin{equation}
\lim_{K,N\rightarrow\infty}\frac{ \La \ln Z(\D) \Ra_\D}{KN} = \min_q 
\half \frac{q}{1-q} + \half \ln(1-q) +
       \alpha \La \ln H(-z_0 \sqrt{ q^{\rm e}}/\sqrt{1-q^{\rm e}}\,)  \Ra_{z_0}
\label{RScom}
\end{equation}
where again the only difference to the corresponding expression for the 
perceptron (\ref{Glog}) is the substitution in the energy term of $q$ by
$q^{\rm e} = \frac{2}{\pi} \arcsin q$. This, however, has a drastic effect
on the capacity, since the derivative of $\arcsin q$ is singular at $q=1$.
As a consequence the energy term,
$\La \ln H(-z_0 \sqrt{ q^{\rm e}}/\sqrt{1-q^{\rm e}}\,)  \Ra_{z_0}$, 
diverges as $1/\sqrt{1-q}$ in the limit $q\rightarrow 1$ instead of
the the $1/(1-q)$ divergence found for the perceptron. Now the divergence
of the entropy term in (\ref{RScom}) for $q\rightarrow 1$ is no longer balanced
by the divergence of the energy term. Hence the minimization problem has 
a solution for all values of $\alpha$ and in particular $1-q$ scales as
$1/\alpha^2$ for large $\alpha$.

So we have found the important result that the storage capacity of the TCM
diverges with the number of hidden units $K$, and in this sense the multilayer
perceptron is more powerful than the sum of its parts. Unfortunately the
calculation yields no information on how quickly the capacity increases
with $K$. 

To gain some insight into this question let us consider the
accuracy of the Gaussian approximation leading to the above result. Going back
to Eq. (\ref{mrep}) we set 
$S_k = (X^1_k({\bf Q}),X^2_k({\bf Q}),\ldots,X^n_k({\bf Q}))^T$.
Then one can show that $G= K^{-1/2}\sum_{k=1}^K S_k$ converges to a Gaussian
by calculating the characteristic function 
$\La e^{ {\rm i} V^T G} \Ra_{ \{S_k\} }$, where $V\in\R^K$. 
This yields 
\begin{eqnarray}
\La e^{ {\rm i} V^T G} \Ra_{ \{S_k\} }
&=& \La e^{ {\rm i} V^T S_1/\sqrt K} \Ra_{S_1}^K \nonumber \\\
&=& \left(1 - \frac{ \La (V^T S_1)^2 \Ra_{S_1} }{2K} + 
    \La {\cal O}\left( \frac{(V^T S_1)^4}{K^2}\right) \Ra_{S_1}\right )^K 
\label{expans}
\end{eqnarray} 
as the odd terms in the expansion vanish because $S_1$ and $-S_1$ have the
same distribution. 
One then argues that the higher order term can be neglected for large $K$, 
and this yields that the characteristic function converges to
\begin{equation*}
e^{- \half \La (V^T S_1)^2 \Ra_{S_1}}  = 
e^{- \half  V^T \La S_1 S_1^T \Ra_{S_1} V } = 
e^{- \half  V^T Q^{\rm e} V}\,,
\end{equation*} 
and this,  being Gaussian, is the characteristic function of a Gaussian.
But now assume that the matrix $Q^{\rm e}$ is close to singular, 
let $\lambda$ be its smallest eigenvalue and $V$ an eigenvector to $\lambda$.
Then in the second term of the expansion (\ref{expans})
the average $\La (V^T S_1)^2 \Ra_{S_1}$ is on the order of $\lambda$ and
quite small and the quadratic term only give the leading correction if
$\lambda$ is large compared to $1/K$, that is if $\lambda K \gg 1$. 
Otherwise, it only make sense to truncate the expansion
after the constant term, 
in essence equating $\lambda$ with zero, or to take the term of 
higher than quadratic order into account as well.

In the replica symmetric theory the smallest eigenvalue of $ Q^{\rm e}$ is
$1-q^{\rm e}$ which approaches zero with increasing $\alpha$. It is impossible
to equate $1-q^{\rm e}$ with zero, since this leads to a divergence of the
energy term in Eq. (\ref{RScom}). So, since the Gaussian approximation ignores
the higher than quadratic  terms, it  can
only be trusted if $(1-q^{\rm e}) K$ is large, and the scaling of $q$ with
$\alpha$ yields that this requires $\alpha \ll \sqrt{K}$.

On the other hand, if  $\alpha \ll \sqrt{K}$, the Gaussian approximation is
reliable and thus the theory predicts that the capacity of the TCM is at least
$\sqrt{K}$. Indeed, using a replica symmetric parameterization of Eq.
(\ref{mrep}) and taking the $n\rightarrow 0$ limit before the large $K$ limit,
has been shown to yield a $\sqrt{K}$ divergence of the capacity 
\cite{Bar92,Eng92}.
Unfortunately these results are completely {\em wrong} as already noticed
in \cite{Bar92,Eng92}. In \cite{Mit89}  
rigorous upper bounds on the capacity are derived, which show that the
storage capacity of the TCM cannot diverge with $K$ faster than $\log K$.

\section{Breaking replica symmetry } \label{RSBG}

It turns out that one does not obtain the correct analytical continuation
from integer $n$ to $n$ close to zero when using a replica symmetric 
parameterization of $Q$. Such a phenomenon was first discovered in the
quite different context of the infinite range spin glass \cite{Mez87} and,
after much soul searching among the involved physicist, Giorgio Parisi
came up with a hierarchical scheme for relaxing the replica symmetric
assumption. I shall first apply the first level of this scheme to the TCM
(one step of replica symmetry breaking or just RSB1) and then discuss the
physical implications of the approach.

The basic idea in RSB1 is to partition the $n$ replicas into $n/m$ groups of 
equal size and parameterize $Q$ by setting $Q^{ab}$ equal to $q_1$ if the
different replicas $a$ and $b$ belong to the same group and to $q_0$ else.
Formally this amounts to writing $Q$ as the block matrix
\begin{equation}
Q = M_{n/m}( M_m(1,q_1), M_m(q_0,q_0) ). \label{RSB1mat}
\end{equation}
It is simple to calculate
the determinant of $Q$ by applying Eq. (\ref{decomp}) to obtain 
\begin{equation*}
\det Q = \det  M_m\left(1+\frac{n-m}{m}q_0 ,q_1+\frac{n-m}{m}q_0 \right)
         \det M_m(1-q_0,q_1-q_0)^{\frac{n-m}{m}}\;.
\end{equation*}

Our next goal is to decompose the $n$-dimensional Gaussian $Y(Q^{\rm e})$
in Eq. (\ref{TCMn}). Note that $Q^{\rm e}$ has the same structure as 
$Q$ with the $q_i$ replaced by $q^{\rm e}_i = \frac{2}{\pi} \arcsin q_i$.
Because of the partitioning of the replicas,
it is convenient to think of the replica index $a$ as a two dimensional index
$a=[u,v]$. 
We define $[u,v] = (u-1)m + v$ for $v = 1,\ldots m$, and 
$u = 1,\ldots,n/m$ indexes the different groups of the partition.
One can now rewrite 
$Y^a = Y^{[u,v]}$ in terms of i.i.d. {\cal N}(0,1) random variables
$z, z^u, z^{u,v}$ as
\begin{equation*}
Y^{[u,v]} = \sqrt{q^{\rm e}_0}\,z + \sqrt{q^{\rm e}_1 -q^{\rm e}_0}\,z^u +
             \sqrt{1 -q^{\rm e}_1}\, z^{u,v}\;.
\end{equation*}
So for the $Y(Q^{\rm e})$ average in the energy term of (\ref{TCMn}) we obtain
\begin{eqnarray*}
\lefteqn{
  \La  \prod_{a=1}^n F(\sign( Y_a(Q^{\rm e})) ) \Ra_{Y(Q^{\rm e})} 
} \\
&=& \La \prod_u \prod_v
        F(\sign(\sqrt{q^{\rm e}_0}\,z + \sqrt{q^{\rm e}_1 -q^{\rm e}_0}\,z^u +
             \sqrt{1 -q^{\rm e}_1}\, z^{u,v})) 
    \Ra_{z^{u,v},z^u,z} \\
&=& \La \La  \La 
       F(\sign(\sqrt{q^{\rm e}_0}\,z + \sqrt{q^{\rm e}_1 -q^{\rm e}_0}\,z^1 +
             \sqrt{1 -q^{\rm e}_1}\, z^{1,1})) 
    \Ra_{z^{1,1}}^{m} \Ra_{z^1}^{n/m} \Ra_z
\end{eqnarray*}
where the last expression makes sense for noninteger $n$ and $m$.
So, using this continuation to small $n$ and assuming that $F=\Theta$,
from Eq. (\ref{TCMn}) we obtain within the RSB1 Ansatz 
\begin{equation}
\lim_{K,N\rightarrow\infty}\frac{ \La \ln Z(\D) \Ra_\D}{KN} =
  \min_{q_0,q_1,m}   G_s(q_0,q_1,m) + G_r(q_0,q_1,m) \;,
   \label{RSBcom}
\end{equation}
where
\begin{eqnarray*}
  G_s &=& \frac{1}{2} \frac{q_0}{1-q_1 +m(q_1-q_0)} + 
                \frac{m-1}{2m}\ln(1-q_1) + \frac{1}{2m}\ln(1-q_1 +m(q_1-q_0))
      \;,\\
   G_r &=& \frac{\alpha}{m} \La
            \ln \La   
            H\left(
             \frac{ \sqrt{q^{\rm e}_0}z - \sqrt{ q^{\rm e}_1-q^{\rm e}_0 }z^1 }
                  { \sqrt{1-q^{\rm e}_1} } 
             \right)^m \Ra_{z^{1}} \Ra_z
                .
\end{eqnarray*}

Let us first discuss the ways in which the RSB1 parameterization reduces to
the replica symmetric one. The case $q_1 = q_0 = q$ is obvious, then $G_s+G_r$
no longer depends on $m$ and is the same as (\ref{RScom}). But also for $m=1$
one finds that $G_s+G_r$ is now independent of $q_1$ and equivalent to the 
replica symmetric expression with $q_0$ playing the r\^ole of $q$. While one
cannot set $m = 0$, a little algebra shows that in the limit
$m\rightarrow 0$ the value of $G_s+G_r$ becomes independent of $q_0$ yielding  
equivalence to the  replica symmetric case with $q_1$ playing 
the r\^ole of $q$.

To solve Eq. (\ref{RSBcom}), it helps to first consider the simpler
problem of minimizing
\begin{equation*}
F(q_0,q_1,m) =  G_s(q_0,q_1,m) + G_r(q_0,q_1,m) 
                  - \frac{m-1}{2m}\ln(1-q_1)-\frac{1}{2m}\ln\, m
\end{equation*}
Setting $q_1=1$, one immediately sees that $m F(q,1,m)$ is independent of $m$
 and equal to replica symmetric functional (\ref{RScom}). 
Since for any $\alpha > 0$
the latter can be made negative by an appropriate choice of $q$, 
for this choice of $q$ the value of $F(q,1,m)$ diverges to $-\infty$ for
$m\rightarrow +0$. So minimizing $F(q_0,q_1,m)$ is easy.

Due to the divergent
$\ln(1-q_1)$ term in $G_s$, one cannot set $q_1=1$ in the function
 (\ref{RSBcom}) we actually want to minimize. 
But since the divergence is only logarithmic,
one will expect the optimal values of $1-q_1$ and $m$ to be close to $0$. 
Using the
asymptotic expansion of $H(x)$ for large arguments, makes it possible
to simplify the energy term for $q_1\rightarrow 1$. In the end one finds
that the minimization problem has a finite solution for all values of $\alpha$
and that $q_0\rightarrow q_1\rightarrow 1$ and $m\rightarrow 0$ with
increasing $\alpha$. The asymptotic scalings are
\begin{equation}
1-q_0 \propto \alpha^{-2}\quad\mbox{and}\quad
\ln(1-q_1) \propto \ln\, m \propto \alpha^2\;. \label{RSBscales}
\end{equation}
In spite of the fact that $q_0\rightarrow q_1$ as well as
$m\rightarrow 0$ are replica symmetric limits, the prediction for the typical
volume is completely different. A super-exponential decay is found in
RSB1, in contrast to the exponential decay of the volume in the replica 
symmetric theory.

Most importantly, we obtain a completely different result for the validity of
the Gaussian approximation. The smallest Eigenvalue $\lambda$ of $Q^{\rm e}$ is
now $1-q^{\rm e}_1$. So  in view of (\ref{RSBscales}) the $\lambda K \gg 1$
criterion for trusting the central limit theorem translates into 
$\alpha \ll \sqrt{\ln K}$ and the Gaussian theory now predicts that the
critical capacity is at least on the order of $\sqrt{\ln K}$ which is 
entirely compatible with the rigorous $\ln K$ upper bound.

\section{The physical meaning of RSB}

To lighten the notation, I shall discuss the interpretation of RSB
in the context of perceptron learning. Since for the perceptron replica
symmetry is broken only beyond capacity, the discussion is based on
the Gibbs density $p(\sigma_J)$ defined by Eq. (\ref{Temp}). One can
then consider the probability density $P_\D(q)$ that the weight vectors 
$J^1$ and $J^2$  of two perceptrons drawn from the Gibbs density have an
overlap $q$. The cumulative distribution function of $P_\D(q)$ is
\begin{eqnarray*}
C_\D(q) &=& \int_{-1}^q {\rm d}x P_\D(x)  \\
&=& 
\int {\rm d}J^1 {\rm d}J^2 \Theta(q-{J^1}^T J^2)p(\sigma_{J^1})p(\sigma_{J^2})
\\
&=&
Z(\D)^{-2}
\int {\rm d}J^1 {\rm d}J^2 \Theta(q-{J^1}^T J^2)
\prod_{a=1}^2\prod_{\mu=1}^P F(\tau^\mu {J^a}^T\xi^\mu)
\end{eqnarray*}
We want to calculate the training set average of $C_\D(q)$ for large $N$ and 
to this end  consider the related quantity
\begin{equation*}
C_\epsilon(q,N,n) = 
\La Z(\D)^{n-2}
\int {\rm d}J^1 {\rm d}J^2 \Theta_\epsilon(q-{J^1}^T J^2)
\prod_{a=1}^2\prod_{\mu=1}^P F(\tau^\mu {J^a}^T\xi^\mu) \Ra_\D
\end{equation*}
where $\Theta_\epsilon(x) = \epsilon + \Theta(x)$. Setting $\epsilon = 0$,
for the object of interest to us we have 
\begin{equation}
\La C_\D(q) \Ra_\D = C_0(q,N,0)\,. \label{sillyrel}
\end{equation}  
But in the sequel we assume that $\epsilon$ is positive, taking the limit 
$\epsilon\rightarrow 0$ in the end.

For integer $n \geq 2$ one has
\begin{equation*}
C_\epsilon(q,N,n) = 
\La
\int {\rm d}\J\,\Theta_\epsilon(q-{J^1}^T J^2)
\prod_{a=1}^n\prod_{\mu=1}^P F(\tau^\mu {J^a}^T\xi^\mu) \Ra_\D.
\end{equation*}
and, using replicas, we evaluate the RHS  for general $n$. To get rid
of the special treatment of the first two indeces, we introduce the function
\begin{equation*}
G_\epsilon(q,Q) = \frac{1}{n(n-1)} \sum_{a\neq b} \Theta_\epsilon(q -Q^{ab})
\end{equation*}
and by symmetry
\begin{equation*}
C_\epsilon(q,N,n) = \La \int {\rm d}\J\, G_\epsilon(q,\J^T\J) 
\prod_{a=1}^n\prod_{\mu=1}^P F(\tau^\mu {J^a}^T\xi^\mu) \Ra_\D.
\end{equation*}
Transforming the integral to the order parameter matrix $Q$ yields
\begin{equation*}
C_\epsilon(q,N,n) =  
D_n({\mathsf 1}) \int \!\dd{Q}\, G_\epsilon(q,Q) 
\left(  (\det Q)^\frac{1-(n+1)/N}{2}
\La  \prod_{a=1}^n F( X_a(Q) ) \Ra_{\!\!X(Q)}^\alpha 
\right )^N .
\end{equation*}
The integral will decay to zero with increasing $N$, and 
the asymptotic rate of decay is to leading order found by
Laplace's method \cite{Cop65}. This shows that 
the decay is determined by the properties
of the integrand in arbitrarily small neighborhoods of its maxima.
In fact, since $\epsilon$ is positive, $G_\epsilon(q,Q)$ does not vanish,
and we need only the neighborhood of a maximum 
 $Q_*(n)$ of
$(\det Q)^\frac{1}{2} \La  \prod_{a=1}^n F( X_a(Q) ) \Ra_{\!\!X(Q)}^\alpha$
if this maximum is unique up to permutations of the replica indeces.
Further $G_\epsilon(q,Q)$ is piecewise constant, and in particular
$G_\epsilon(q,Q) = G_\epsilon(q,Q_*(n))$ holds in a neighborhood of the
maximum, except if $q$ is equal to an off-diagonal element of $Q_*(n)$.
So for a generic value of $q$ we can treat $G_\epsilon(q,Q)$ as a factor
constant in $Q$ and find
\begin{equation*}
\lim_{N\rightarrow\infty} 
 \frac{C_\epsilon(q,N,n)}{C_\epsilon(1,N,n)} = 
 \frac{G_\epsilon(q,Q_*(n))}{G_\epsilon(1,Q_*(n))}\;.
\end{equation*}
Thus in any  parameterization of $Q_*(n)$ which enables us to take $n$ to zero,
considering the limit $\epsilon\rightarrow 0$, we obtain for the cumulative 
distribution of $q$
\begin{equation*}
\lim_{N\rightarrow\infty} \La C_\D(q) \Ra_\D = 
\lim_{n\rightarrow 0} G_0(q,Q_*(n))\,.
\end{equation*}
We have used (\ref{sillyrel}) and the fact that 
$G_0(1,Q_*(n)) = 1$.

In replica symmetry this yields the simple result that 
$\La C_\D(q) \Ra_\D$ approaches the step function $\Theta(q-q_*)$ for large 
$N$, where $q^*$ is the stationary value of the order parameter for 
small $n$. Further, since $\La C_\D(q) \Ra_\D$ converges to a step function 
with a single step, $C_\D(q)$ is selfaveraging.

If $Q$ is the RSB1 matrix (\ref{RSB1mat}), one has 
$G_0(q,Q) = \frac{n-m}{n-1} \Theta(q-q_0) +  \frac{m-1}{n-1} \Theta(q-q_1)$.
So setting $n$ to zero we obtain as the physical interpretation of RSB1 that
the cumulative distribution of $q$ has two steps:
\begin{equation*}
\lim_{N\rightarrow\infty} \La C_\D(q) \Ra_\D =
m \Theta(q-q_0) + (1-m)\Theta(q-q_1)\;. 
\end{equation*}
However, when replica symmetry is broken, $C_\D(q)$ is no longer 
selfaveraging \cite{Mez87,Gyo01}. So, even for large $N$, not all properties
of the single system, can be deduced from those of the training set
average in this case.

\section{Beyond RSB1 ?}

For the TCM discussed in Section 2 the order parameter $q$ refers to the
overlap between the weight vectors of the same hidden unit of two TCM's in 
version space. So, the RSB1 solution, means that the pattern averaged
density of this overlap converges to two $\delta$-peaks. Why not three?

Indeed, it is straightforward two allow for three peaks by parameterizing
$Q$ as
\begin{equation}
Q = M_{n/m_1}( R_1(m_1,m_2;1,q_2,q_1),R_1(m_1,m_2;q_0,q_0,q_0)),
\end{equation}
where $R_1$ denotes an RSB1-matrix,
$
 R_1(n,m;a,b,c) = M_{n/m}( M_m(a,b), M_m(c,c) )
$.  Of course, one might still think that this RSB2 Ansatz is not general
enough, and recursively continue to construct an RSB-$k$ parameterization
allowing for $k+1$ peaks.

Using the techniques discussed for RSB1, it is straightforward, if somewhat 
tedious, to write down the minimization problem for the typical volume using
a, say, RSB2 parameterization. And I would expect, that for sufficiently large
but finite $\alpha$ a higher order RSB parameterization does improve on RSB1.
But it is not clear that this will affect the capacity result,
because the higher order solutions can converge to the RSB1 solution with
increasing $\alpha$. One will in fact expect the $q$'s to converge 
to $1$, that is toward a single peak, and already in the RSB1  
parameterization, rather extreme scalings are needed to construct a non 
replica symmetric solution. Unfortunately, it would be extremely complicated,
to show analytically, that any RSB2 solution must be
degenerate with the RSB1 solution (\ref{RSBscales}) in the large 
$\alpha$ limit. Indeed I have not even proven that the solution 
(\ref{RSBscales}) is the
unique global minimum in RSB1 space. But it would perhaps be worthwhile to
numerically track a higher order RSB solution to large values of $\alpha$.

%% file: mlp2.tex
\section{Storage capacity of the CM}

We now obtain an accurate value for the capacity of the CM by 
taking the limit $n\rightarrow 0$ limit for finite $K$. 
Going back to Eq. (\ref{mrep}) and using the site symmetry
assumption (\ref{sitesym}) we consider the RSB1-theory.
Since we are dealing with the fully connected architecture,
the matrices $Q$ and $P$ are parameterized as:
\begin{equation}
Q = M_{n/m}( M_m(1-\frac{p_2}{K},q_1), M_m(q_0,q_0) ),\quad
P = M_{n/m}( M_m(  p_2  ,p_1), M_m(p_0,p_0) )\,.  \label{RSB1par}
\end{equation}
To decompose the Gaussians $X^a_k({\bf Q})$ in Eq. (\ref{mrep}),
we rewrite $a$ in form of the two dimensional index $[u,v]$ as in Section 
\ref{RSBG}, employ the i.i.d. ${\cal N}(0,1)$ Gaussians
$z_k,z_k^u,z_k^{u,v}$ and set:
\begin{equation}
X^{[u,v]}_k = u z_k + v z_k^u + w z_k^{u,v} + 
        \bar u \bar z + \bar v \bar z^u + \bar w \bar z^{u,v}. \label{decomp1}
\end{equation} 
Here the parameters ($u,v$ etc.) have to be chosen so that 
${\bf Q} = M_K(Q+P/K,P/K)$ holds for the 
covariance matrix $\bf Q$ of the $X^a_k$. The last three summands in 
(\ref{decomp1}) are needed because the sites can be correlated 
$(P\neq 0)$. In keeping with our usual style of decomposing Gaussians
one might expect $\bar z, \bar z^u,\bar z^{u,v}$ to be ${\cal N}(0,1)$ 
Gaussians independent of each other and the other random variables. 
However, it turns out that in the relevant regime the sites are 
anti-correlated,
$\La X^a_k X^a_l \Ra < 0$ for $k\neq l$. If all random variables
in the decomposition are independent, such anti-correlations are only possible
if, say, $\bar u$ is imaginary. This is probably not really a problem because
the averages in the energy term lead to $H$-functions which do make sense
for complex arguments. However, I find this too murky, $X^{[u,v]}_k$
is after all a real valued Gaussian, and thus adopt a different definition of 
$\bar z, \bar z^u,\bar z^{u,v}$, setting:
\begin{equation*}
\bar z = K^{-1}\sum_{k=1}^K z_k,\quad 
\bar z^u = K^{-1}\sum_{k=1}^K z^u, \quad 
\bar z^{u,v} = K^{-1}\sum_{k=1}^K z^{u,v}_k. 
\end{equation*}
Then a simple calculation shows, that the $X^a_k$ have the desired covariances
if the parameters satisfy
 \begin{equation}
\begin{array}[b]{lclllcl}
\# u^2 &\#=&\# q_0         &\#&\# (u+\bar{u})^2 &\#=&\# p_0 + q_0\\
\# v^2 &\#=&\# q_1-q_0     &\#&\# (v+\bar{v})^2 &\#=&\# q_1+p_1 - q_0-p_0\\
\# w^2 &\#=&\# 1-p_2/K-q_1 &\#&\# (w+\bar{w})^2 &\#=&\# 1-p_2/K+p_2 -q_1-p_1 
\label{oparrels}
\end{array}.    
\end{equation}
Using this decomposition the average in the energy term of Eq. (\ref{mrep})
can be rewritten as:
\begin{eqnarray*}
\lefteqn{
\La \prod_{a=1}^n F\left(\sign(\sum_{k=1}^K \sign(X^a_k({\bf Q}))) 
           \right) \Ra_{X({\bf Q})}} \\
&=&\La \La  \La 
       F\left(\sign(\sum_{k=1}^K\sign( u z_k + v z_k^1 + w z_k^{1,1} +
      \bar u \bar z + \bar v \bar z^1 + \bar w \bar z^{1,1} ))\right) 
    \Ra_{\!\!\{z_k^{1,1}\}}^{\!\!m} \Ra_{\!\!\{z_k^1\}}^{\!\!n/m} 
    \Ra_{\!\!\{z_k\}} \\
&=&\La \La  \La 
       F\left(\sign(\sum_{k=1}^K\sign(X_k^{[1,1]} ))\right) 
    \Ra_{\!\!\{z_k^{1,1}\}}^{\!\!m} \Ra_{\!\!\{z_k^1\}}^{\!\!n/m} 
    \Ra_{\!\!\{z_k\}}\,.
\end{eqnarray*}
To lighten the notation in the last equation (\ref{decomp1}) is used to write
the fields in a more compact form.

For the entropy term, we need $\det\!\bf Q$ and the determinant is easily 
calculated by repeatedly applying (\ref{decomp}). Specializing to $F=\Theta$,
from the small $n$ limit one then obtains:
\begin{equation}
\lim_{N\rightarrow\infty}\frac{ \La \ln Z(\D) \Ra_\D}{KN} =
  \exta{\{q_i\},\{p_i\},m} \frac{1}{m}G_s(\{q_i\},\{p_i\},m) + 
                    \frac{\alpha}{m} G_r(\{q_i\},\{p_i\},m)
   \label{RSBKcom}
\end{equation}
where 
\begin{eqnarray*}
G_s &=& \frac{K-1}{2K}S\left(u^2,v^2,w^2,m\right) + 
      \frac{1}{2K}S\left((u+\bar{u})^2,(v+\bar{v})^2,(w+\bar{w})^2,m\right)
      \nonumber \\
S(a,b,c,m) &=& (m-1)\ln c + \ln(c+mb) + \frac{ma}{c+mb}\;
\end{eqnarray*}
and
\begin{equation*}
G_r = \La \ln\La  \La 
       \Theta\left(\sign(\sum_{k=1}^K\sign(  X_k^{[1,1]} ))\right) 
    \Ra_{\!\!\{z_k^{1,1}\}}^{\!\!m} \Ra_{\!\!\{z_k^1\}}
    \Ra_{\!\!\{z_k\}}.
\end{equation*}
The extremum in Eq. (\ref{RSBKcom}) means that the function has to be
minimized w.r.t to all order parameters except $p_0$. The function  has to
be maximized w.r.t to $p_0$ since in physical terms 
$p_0 = K {J_k^a}^T J_k^a$. So $p_0$ refers to a quantity with a single replica
index and is analogous to the student/teacher overlap in the context of
learning a rule in the following sense: Instead of considering the full space
of networks, we could have focused on sub-shells where $J_k^T J_k$ is 
constant. The capacity of the full space of networks is given by the sub-shell
of maximal capacity and this is just what is obtained by maximizing Eq. 
(\ref{RSBKcom}) in $p_0$.

It is impractical to calculate $G_r$ in its present form since this involves 
$3K$ Gaussian  integrals. To do something about this, we first rewrite 
(\ref{RSBKcom}) in terms of the internal representations $\iota$ 
in the committee, that is in terms 
of the outputs $\iota_k$ of its hidden units. We use 
\begin{equation*}
1 = {\rm Tr}_\iota \prod_{k=1}^K \Theta(\iota_k X_k^{[1,1]})\,,
\end{equation*}
where the trace over $\iota$ denotes a summation over all 
$\iota\in\{-1,1\}^K$. 
Multiplying the $\Theta$-function in $G_r$ with the above RHS yields:
\begin{equation*}
G_r = \La \ln\La \La 
        {\rm Tr}_\iota  \Theta\left(\sum_{k=1}^K \iota_k\right)
        \prod_{k=1}^K \Theta(\iota_k X_k^{[1,1]})
      \Ra_{\!\!\{z_k^{1,1}\}}^{\!\!m} \Ra_{\!\!\{z_k^1\}}
    \Ra_{\!\!\{z_k\}}.
\end{equation*}
To highlight the dependence of the energy term on $w$ and $\bar w$,
we now define:
\begin{equation}
f(\{Y_k\},\{\iota_k\}) =
\left\langle\prod_{k=1}^K \Theta[\iota_k(Y_j+w z^{1,1}_k+\bar{w}\bar z^{1,1})]
\right\rangle_{\{z^{1,1}_k\}}
\mbox{\ and\ }
Y_k = u z_k+\bar{u}\bar{z}+v z^1_k+\bar{v} \bar z^1\;,
\label{fdef}
\end{equation}
so
\begin{equation*}
G_r =  \La \ln\La  
        \left({\rm Tr}_\iota  \Theta\left(\sum_{k=1}^K \iota_k\right)
        f(\{Y_k\},\{\iota_k\})\right)^m \Ra_{\!\!\{z_k^1\}}
    \Ra_{\!\!\{z_k\}}.
\end{equation*}
As $\alpha$ approaches the critical capacity the volume of admissible 
networks vanishes and one will expect that $w,\bar w \rightarrow 0$.
In this limit the trace is dominated by a single term and thus
\begin{equation*}
 \left({\rm Tr}_{\iota} 
        \Theta\left( {\sum_k \iota_k} \right)f(\{Y_k\},\{\iota_k\})\right)^m
\sim
   \max_{\iota} 
        \Theta\left( {\sum_k \iota_k}\right)f(\{Y_k\},\{\iota_k\})^m\,.
\end{equation*}
It is too troublesome to locate the maximum as function of the $Y_k$ and
hence we replace the maximization in $\iota$ by a summation over all possible
values of $\iota$. So we use that for $w,\bar w \rightarrow 0$ 
\begin{equation}
 \left({\rm Tr}_{\iota} 
        \Theta\left( {\sum_k \iota_k} \right)f(\{Y_k\},\{\iota_k\})\right)^m
\leq
   {\rm Tr}_{\iota} 
        \Theta( {\displaystyle\sum_k \iota_k} )f(\{Y_k\},\{\iota_k\})^m .
\label{ineq}
\end{equation}
The simplified energy term 
\begin{equation*}
\hat G_r = \La \ln\La  
        {\rm Tr}_\iota  (\Theta\left(\sum_{k=1}^K \iota_k\right)
        f(\{Y_k\},\{\iota_k\}))^m \Ra_{\!\!\{z_k^1\}}
    \Ra_{\!\!\{z_k\}}
\end{equation*}
obtained by commuting the trace and the exponentiation with $m$ is an
upper bound to the true value of $G_r$ for $w,\bar w \rightarrow 0$, and
we consider the extremal problem
\begin{equation}
  \exta{\{q_i\},\{p_i\},m} G_s(\{q_i\},\{p_i\},m) + 
                    \alpha \hat G_r(\{q_i\},\{p_i\},m)
   \label{RSBKcoms}
\end{equation}
instead of (\ref{RSBKcom}). 

While (\ref{RSBKcoms}) is more accessible than the original extremal
problem, the remaining calculations are nevertheless
quite involved and they are described in some detail in \urbseven. 
Here, I shall just present the key features of the solution. At a critical 
value  $\hat\alpha_c(K)$ of $\alpha$  which for large $K$ scales
as 
\begin{equation*}
\hat\alpha_c(K) \sim \frac{16}{\pi-2}\sqrt{\ln K}\ 
\end{equation*}
the solution of (\ref{RSBKcoms}) diverges. At the critical $\alpha$
the stationary values of $w$ and $\bar w$ vanish. Consequently
the extremal value of (\ref{RSBKcoms}) bounds the extremal
value of the original problem (\ref{RSBKcom}) and $\hat\alpha_c(K)$ is
an upper bound to the true capacity $\alpha_c(K)$. An interesting question is,
whether this upper bound coincides with the critical capacity to leading 
order in $K$.  This is related to the question if the inequality (\ref{ineq})
is sufficiently tight at the stationary point. 
Since the trace over the internal representations $\iota$ on its 
LHS is dominated by a single term due to $w,\bar w \rightarrow 0$, the 
inequality would be tight if also on the RHS the trace were dominated by a
single term. This, however, is tricky, since as $\alpha$ approaches 
$\hat\alpha_c(K)$ one finds that also $m\rightarrow 0$. 
Consequently as function
of $\iota$ the maximum of $f(\{Y_k\},\{\iota_k\})^m$ on the RHS is not as
pronounced as the one of just $f(\{Y_k\},\{\iota_k\})$ on the LHS of the 
inequality. So the tightness of the inequality is determined by the ratios of 
$w,\bar w$ and $m$ as they approach zero. 
These are calculated in \urbseven\ and suggest 
that while a few terms do contribute to the trace on the RHS, their number is
not very large, and that the critical capacity to leading order coincides
with $\hat \alpha_c(K)$. It would however require quite intricate 
combinatorics to actually show that this is the case.

Finally, it is interesting to compare the results for the connected committee
to the ones for the tree architecture. For the TCM entirely analogous 
calculations \cite{Mon95,Mon96} yield the smaller capacity of
$\frac{16}{\pi}\sqrt{\ln K}$. This difference is due to the fact that at
the critical capacity the weight vectors of the CM are anti-correlated, 
$p_0 = -1$. If one were to artificially restrict the state space of the CM 
so that the $K$ weight vectors are forced to be orthogonal, 
this corresponds to $p_0 = 0$, the capacities of the CM and the TCM would be 
the same. The usefulness of anti-correlated hidden units is related to the
fact that in the orthogonal case the output of the CM is quite similar to that
of a perceptron. In particular, if one considers the perceptron with weights
$\bar J$ obtained by averaging the weight vectors $J_k$ of the CM, 
$\bar J = \sum_{k=1}^K J_k$, one finds that when $p_0=0$ this perceptron 
gives the same output as the CM for approximately 80\% of randomly chosen 
inputs even when $K$ is large. However, $p_0=-1$ leads to $\bar J = 0$,
the approximating perceptron is undefined, and no perceptron improves on
random guessing in predicting the output of this CM for large $K$. 
Since the storage 
capacity of the perceptron is limited, the anti-correlated state maximizes
the capacity of the committee.

\section{Counting internal representations}

Historically, the capacity of the committee machine was first obtained by
counting the typical number of internal representation 
of a training set \cite{Mon95,Mon96} and not by the RSB1 calculations 
of the Gardner volume.
To round off the analysis of the CM, I shall describe the close relationship
between the two approaches.

Given a training set $\D =  \{(\xi^{\mu},\tau^\mu)\}$ one can ask whether
outputs $\iota_k^\mu\in \{-1,1\}$ of the hidden units exist which
can be (a) realized by the committee and for which (b) the output of the
committee  on $\xi^\mu$ is $\tau^\mu$. This amounts to asking whether the 
volume of weights
\begin{equation}
V_{\iota}(\D) =
   \prod_{\mu=1}^P 
   \Theta(\tau^\mu {\textstyle \sum_{k=1}^K \iota_k^{\mu}})
   \int \!\dd{\J} \prod_{\mu=1}^P\prod_{k=1}^K 
   \Theta(\iota_k^{\mu} J_k^T \xi^{\mu})\;.  \label{Vtau}
\end{equation}
associated with the internal representation $\iota$ is nonzero.
The are $2^{(K-1)P}$ internal representations with the property (b),
$\tau^u {\textstyle \sum_{k=1}^K \iota_k^{\mu}} > 0$, but not all of
them will be realizable by the committee. So the quantity of interest
is the typical number of realizable representations 
\[
\exp \La \ln {\rm Tr}_\iota \Theta(V_\iota(\D)) \Ra_\D\,.
\]
To obtain the training set average one uses a double replication.
Instead of $\Theta(V_\iota(\D))$ one considers $V_\iota(\D)^m$ for 
integer $m$ taking the limit $m\rightarrow 0$ in the end;  the second
replication is used to calculate the logarithm in the usual way.
We thus consider
\begin{equation}
S(m) = \frac{1}{KN} \frac{ \dd{\ } }{ \dd{\hat n}} 
       \left[\La  \left({\rm Tr}_\iota V_\iota(\D)^m \rule{0em}{1.25em}
                  \right)^{\hat n} \Ra_\D
       \right]_{| \hat n = 0}
\end{equation}
and are mainly interested in $S(0) = \lim_{m\rightarrow 0} S(m)$. As long
as $S(0)$ is positive, realizable internal representations exist, and the
storage capacity of the committee is not exhausted. For $P= \alpha KN$, the
smallest value $\alpha_d(K)$ for which  $S(0) = 0$ marks the transition to
a regime where the number of internal representations is no longer exponential
in $N$. So $\alpha_d(K)$ is a lower bound on the capacity and for finite $K$
one will not expect the bound to be tight; for instance $\alpha_d(1)=0$ but
the critical capacity for $K=1$ is $2$. It is, however, reasonable to expect
that $\alpha_d(K)$ for large $K$ yields an asymptotically tight bound 
since 
the volume in weight space associated with any single internal representation
should vanish in this limit. 

To calculate $S(m)$ we use Eq. (\ref{Vtau}) and obtain in a first step
\begin{equation}
\left({\rm Tr}_\iota V_\iota(\D)^m \rule{0em}{1.25em}
                  \right)^{\hat n} =
\int \!\dd{\J} \prod_{\mu,u}   
     {\rm Tr}_\iota \Theta(\tau^\mu {\textstyle \sum_{k=1}^K \iota_k^{\mu}})
     \prod_{u,v}  \Theta(\iota_k^{\mu} {J_k^{uv}}^T \xi^{\mu})\,,
\label{fstep}
\end{equation}
where the replica index $v$ runs from $1$ to $m$, and for the other replica
index: $u = 1,\ldots,\hat n$. The symbol $\dd{\J}$ refers to
 $K\hat n m$ integrations over unit length weight vectors 
$J_k^{uv}$ in $\R^N$. We now have to perform the training set average which,
after commuting with the weight integral, can be rewritten in
term of zero mean Gaussian random variables $X_k^{uv}({\bf Q})$ with 
covariances
\[
\La X_k^{uv}({\bf Q}) X_{k'}^{u'v'}({\bf Q}) \Ra = {\bf Q}_{kk'}^{uvu'v'}
= {J_k^{uv}}^T J_{k'}^{u'v'}
\] 
Transforming to an integral over the order parameter matrix then yields
\begin{eqnarray*}
\lefteqn{
\lim_{N\rightarrow\infty} \frac{
\La  \left({\rm Tr}_\iota V_\iota(\D)^m \right)^{\hat n} \Ra_\D}{KN} =
} \\ & &
\max_{\bf Q}\,\, 
\alpha\, \ln \La \prod_{u}   
  {\rm Tr}_\iota \Theta({\textstyle \sum_{k=1}^K \iota_k^{\mu}})
  \prod_{u,v}  \Theta(\iota_k^{\mu} X_k^{uv}({\bf Q})) \Ra_{X({\bf Q})}
  \!\!+ \frac{\ln\det{\bf Q}}{2K}\;.
\end{eqnarray*}
To make further progress we need to parameterize $\bf Q$. Referring back to 
Eq. (\ref{fstep}), we see that the 
weight vectors $J_k^{uv}$ and $J_{k'}^{u'v'}$ belong to committee machines
which use the same internal representation to store the training set if
$u = u'$. So, even when aiming for a replica symmetric parameterization,
it makes sense to assume that  ${\bf Q}_{kk'}^{uvu'v'}$ depends on 
$\delta_{uu'}$. Further, to control the number of order parameters, we
need to assume site symmetry, that is ${\bf Q}$ depends on $k$ and $k'$ 
only via
$\delta_{kk'}$. These considerations motivate parameterizing $\bf Q$ as
${\bf Q} = M_K(Q+P/K,P/K)$ where\
\[
Q = M_{\hat n}( M_m(1-\frac{p_2}{K},q_1), M_m(q_0,q_0) ), \quad
P = M_{\hat n}( M_m(  p_2  ,p_1), M_m(p_0,p_0) )\,.
\]
Now, comparing to Eq. (\ref{RSB1par}), we see that this is just the 
RSB1-parameterization used in the Gardner volume calculation for the CM if
we equate  $\hat n = n/m$. So we have already calculate $\det\bf Q$ and
the same decomposition of $X(\bf Q)$ into independent contributions
as in the preceding section can be used. We then obtain the following
remarkable analogy to the calculation of the Gardner volume:
\[
S(m) = 
  \exta{\{q_i\},\{p_i\}} G_s(\{q_i\},\{p_i\},m) + 
                    \alpha  \hat{G}_r(\{q_i\},\{p_i\},m)\,,
\]
where $G_s$ and $\hat{G}_r$ are exactly the same as in the preceeding
section. The critical capacity $\alpha_d(K)$ is given by the condition 
$S(0) = 0$. In the RSB1-calculation the bound $\hat\alpha_c(K)$ was
obtained from the divergence to $-\infty$ of $(G_s+ \hat{G}_r)/m$ when
this expression was also minimized w.r.t. to $m$.  But minimizing in $m$ 
yielded that $m \rightarrow 0$ as $\alpha$ approaches $\hat\alpha_c(K)$ and 
this is just the limit needed when counting internal representations.
So we obtain the simple result that
\[
\alpha_d(K) = \hat\alpha_c(K) \sim \frac{16}{\pi-2}\sqrt{\ln K}.
\]
In addition we now have a very nice interpretation of the order parameters, 
e.g.
the overlaps $q_1,p_2$ refer to networks which use the same internal
representation to store the training patterns, 
whereas networks with differing internal representations yield the overlaps
$q_0,p_1$.

However, all is not well. By definition $\alpha_d(K)$ should be a lower
bound to the critical capacity, but $\hat\alpha_c(K)$ is an upper bound.
This shows that the above (doubly) replica symmetric parameterization is too 
simple minded, and replica symmetry is broken, presumably for networks which 
use different internal representations. However, having argued that both
$\alpha_d(K)$ and $\hat\alpha_c(K)$ are tight bounds in the limit of large
$K$, one can reasonably assume that this complication does not 
invalidate the asymptotic findings. This is supported by results in  
\cite{Mon96} where the stability of the replica symmetric stationary
point was analyzed for the TCM when counting internal representations. 
The replica symmetric 
solution was found unstable for finite values of $K$ but marginally stable
in the large $K$ limit.